\documentclass[10pt,letterpaper]{article}
\usepackage[top=0.85in,left=2.75in,footskip=0.75in]{geometry}

\usepackage{amsmath,amssymb}

\usepackage{changepage}

\usepackage[utf8x]{inputenc}

\usepackage{textcomp,marvosym}

\usepackage{cite}

\usepackage{nameref,hyperref}

\usepackage[right]{lineno}

\usepackage{microtype}
\DisableLigatures[f]{encoding = *, family = * }

\usepackage[table]{xcolor}

\usepackage{array}

\newcolumntype{+}{!{\vrule width 2pt}}

\newlength\savedwidth

\raggedright
\usepackage[aboveskip=1pt,labelfont=bf,labelsep=period,justification=raggedright,singlelinecheck=off]{caption}

\makeatletter
\renewcommand{\@biblabel}[1]{\quad#1.}
\makeatother

\usepackage{lastpage,fancyhdr,graphicx}
\usepackage{epstopdf}
\fancyheadoffset[L]{2.25in}
\fancyfootoffset[L]{2.25in}
\begin{document}
\vspace*{0.2in}

\begin{flushleft}
{\Large
\textbf\newline{Comparison of existing aneurysm models and their path forward} 
}
\newline
\\
John Friesen\textsuperscript{1},
Jonas Bergner\textsuperscript{1\ddag},
Mohammad Ibrahim Aftab Khan\textsuperscript{1\ddag},
Stefan Triess\textsuperscript{1\ddag},
Andreas Zoll\textsuperscript{1\ddag},
Peter F. Pelz\textsuperscript{1},
Farzin Adili\textsuperscript{2},
\\
\bigskip
\textbf{1} Chair of Fluid Systems, Technische Universität Darmstadt, Germany
\\
\textbf{2} Department of Vascular Surgery, Klinikum Darmstadt, Germany
\\
\bigskip

%
%

\ddag These authors contributed equally to this work.

\textcurrency Current Address: Chair of Fluid Systems, Technische Univeristät Darmstadt, Germany 



* john.friesen@fst.tu-darmstadt.de
\end{flushleft}
\section*{Abstract}
The two most important aneurysm types are cerebral aneurysms (CA) and abdominal aortic aneurysms (AAA), accounting together for over 80\% of all fatal aneurysm incidences. To minimise aneurysm related deaths, clinicians require various tools to accurately estimate its rupture risk. For both aneurysm types, the current state-of-the-art tools to evaluate rupture risk are identified and evaluated in terms of clinical applicability. 
We perform a comprehensive literature review, using the Web of Science database. Identified records (3127) are clustered by modelling approach and aneurysm location in a meta-analysis to quantify scientific relevance and to extract modelling patterns and further assessed according to PRISMA guidelines (179 full text screens). Beside general differences and similarities of CA and AAA, we identify and systematically evaluate four major modelling approaches on aneurysm rupture risk:
finite element analysis and computational fluid dynamics as deterministic approaches and machine learning and assessment-tools and dimensionless parameters as stochastic approaches. The latter score highest in the evaluation for their potential as clinical applications for rupture prediction, due to readiness level and user friendliness. Deterministic approaches are less likely to be applied in a clinical environment because of their high model complexity. Because deterministic approaches consider underlying mechanism for aneurysm rupture, they have improved capability to account for unusual patient-specific characteristics, compared to stochastic approaches.
We show that an increased interdisciplinary exchange between specialists can boost comprehension of this disease to design tools for a clinical environment. By combining deterministic and stochastic models, advantages of both approaches can improve accessibility for clinicians and prediction quality for rupture risk.



\section{Introduction}

\label{chap:Introduction}
An aneurysm is a permanent focal dilatation of an artery or vein by at least 50\% of its maximum physiological diameter. Aneurysmal dilatation of a blood vessel causes local weakening of the diseased vascular wall and subsequent rupture with potentially lethal bleeding and death. Ruptured aneurysms lead to about 25000 deaths annually in the UK, Germany and the USA combined, accounting for roughly {0.6}{\%} of all yearly fatalities \cite{WHO_Database.2020}. \par 

In general, aneurysm can form anywhere in the blood vessel system, yet they tend to develop mostly in the arterial system of the human body \cite{Lasheras.2007}. The two most common types of aneurysms are the abdominal aorta aneurysm (AAA) and the cerebral aneurysm (CA). AAAs and CAs combined account for over {80}{\%} of fatal aneurysm incidences \cite{WHO_Database.2020}. While AAA are primarily located in the infrarenal aorta \cite{Humphrey.2008}, CA are found in and around the circle of Willis, which supplies blood to the brain and its surrounding structures \cite{Thubrikar.2007} Besides the different locations, most AAAs are of fusiform shape, whereas more than {90}{\%} of CA are of saccular (spherical or berry-like) shape \cite{Lasheras.2007}. Due to the high relevance, the predominant geometry of each type and the abundance of research literature, the focus of this paper is narrowed down to literature regarding fusiform AAAs and saccular CAs.\par

While AAAs are managed and researched mostly by vascualr surgeons, CAs are primarily diagnosed and treated by neuroradiologists and neurosurgeons \cite{Lasheras.2007}. 
The question arises, whether these types of aneurysms are inherently different, or if they share a similar etiology, pathogenesis, biomechanics, and growth until rupture occurs. While some reviews have already touched upon this topic \cite{Lasheras.2007, Humphrey.2008, Tanweer.2014, Hoskins.2017} there is still a pressing need for a comparative approach, especially considering the large amount of new research published in both domains in recent years. For this reason, this review aims to point out similarities regarding initiation, growth and rupture of AAAs and CAs and to promote interdisciplinary work between the different disciplines (biomechanics, vascular and neurosurgery).\par

Th quest for identifiying and translating the underlying mechanisms for aneurysm formation and growth present an ongoing and much-debated research process. In general, hemodynamics, the dynamics of blood flow, act upon the arterial wall as pressure and shear \cite{Lasheras.2007}. Altered hemodynamic conditions over a period of time cause adaption of the arterial wall by remodeling structure and geometry, which in turn further impair the blood flow. While some aneurysms remain stable, others increase in size and eventually rupture. Rupture of an aneurysm occurs from a mechanical viewpoint, when stress acting on the wall locally exceeds the wall's failure strength \cite{Salman.2019}. Rupture of CA lead to subarachnoid haemorrhage (SAH) with a mortality rate of {45}{\%} \cite{Hoskins.2017}. AAA rupture causes internal bleeding with a mortality between {65}{\%} and {80}{\%} \cite{Humphrey.2008}.\par

Most aneurysms are clinically asymptomatic before rupture \cite{Hoskins.2017}, making preventive measures only possible, if they are coincidentally discovered during routine check-ups or targeted screening programs. The most common imaging method used in screening programs for AAAs is ultrasonography. Further screening methods include contrast-enhanced computed tomography (CT) and magnetic resonance imaging (MRI) \cite{Moll.2011}. For patients with an intact aneurysm, physicians face the challenge to give a recommendation whether to treat the aneurysm with inherent risk or to follow up the patient and act upon significant diameter increases above a defined threshold diameter, but potentially risk sudden rupture. This recommendation is given under the constraints of an evident level of uncertainty due to the limited availability of clinical data and time.\par

The options for procedure are either open or endovascular aneurysm repair (EVAR) \cite{Moll.2011}. For AAAs, EVAR has an about four times lower operative mortality rate than open repair ({1.2}{\%} compared to {4.6}{\%} - {4.8}{\%}) \cite{Prinssen.2004, Schermerhorn.2008}. Despite the initial advantage, after three years the overall mortality rates of EVAR and open repair are about the same \cite{Prinssen.2004, Schermerhorn.2008}. There is a continuous need for both procedures, because of possible unsuitability of EVAR or personal preferences of the patient \cite{Moll.2011}. \par

In case of AAAs, the European Society for Vascular Surgery recommends surgery for a maximum aortic diameter of {5.5}{cm} for men and {5}{cm} for females and patients with increased probability of rupture, i.e. smokers and patients with hypertension or chronic airways disease \cite{Moll.2011}. As diameter they recommend using the external aortic diameter in the anterior-posterior plane \cite{Moll.2011}, because of a suggested higher repeatability than the transverse diameter for ultrasonography \cite{Ellis.1991}. The cut-off diameter criterion for an AAA is considered to be insufficient for such a complex disease \cite{Vorp.2007,Kontopodis.2016}. While {13}{\%} of AAAs with a diameter smaller than {5}{cm} rupture, {60}{\%} of AAAs with a diameter greater than {5}{cm} stay intact \cite{Darling.1977}. Furthermore, the use of outdated medical imaging technology, in the studies from which the diameter criterion originates (UKSAT \cite{Powell.1996} and ADAM \cite{Lederle.2000}), leave room for doubt about its accuracy \cite{Kontopodis.2016}. Besides the diameter, the expansion rate of the aneurysm is another important rupture risk factor. For an AAA rapid growth with an expansion rate of $>{1}${cm per year} is seen as critical \cite{Moll.2011}. But due to the need of historical patient data, the expansion rate may not be at hand for a clinical assessment \cite{Bengtsson.1993}. \par

Looking at CAs, the annual risk of rupture is lower ($<{1}${\%} p.a.) compared to that of AAAs (between {3}{\%} p.a - {9}{\%} p.a.) \cite{Wiebers.2003, Chalouhi.2013, Humphrey.2008}. Since CAs are inherently more stable, assessing the rupture risk for an individual case remains a great challenge. While the diameter of the aneurysm is also one of the most commonly used parameters for rupture risk assessment, it is not as clear-cut as for AAAs. The difference between average ruptured to intact aneurysm is {1.5}{mm} at best and over {75}{\%} of ruptured aneurysms are below {10}{mm} in size \cite{Orz.2015, Wiebers.1998, Joo.2009}. For that matter, the American Heart Association lists age, location of the aneurysm, sex, hypertension, smoking, family history and other factors, which have to be taken into account for rupture risk assessment \cite{Thompson.2015}. Still, even after considering these factors, there still remains a high degree of uncertainty regarding the risk of rupture.\par

Various approaches of computational and statistical modelling show potential in a search for a more accurate evaluation of aneurysm rupture probability in a clinical environment. Those approaches are based on computational fluid dynamics (CFD), finite element analysis (FEA), machine learning (ML), assessment-tools \& dimensionless parameters (AT\&DP) and combinations of the above mentioned. The large quantity and variety of approaches relates to the high relevance of aneurysm research. For physicians and researchers alike, it is challenging to keep up, not only with new concepts of their own field, but also with methods that approach aneurysm rupture risk assessment from a different point of view. \\
Therefore, we aim to answer the following two research questions:
\begin{enumerate}
    \item What are similarities and differences of cerebral and abdominal aortic aneurysms regarding their etiology, growth and rupture?
    \item Given the latest state of the art of the modelling approaches to predict aneurysm rupture probability based on a systematic literature search what are open research questions, future directions of each approach and what consequently seems as a promising tool for clinical application? 
\end{enumerate}
Section \ref{sec:ComparisonAAA-CA} focuses upon the first research question, starting with the formation of aneurysms and following the development. At the start of \ref{sec:SystematicModelReview} a brief overview over the different modelling approaches is given, and the search strategy is explained. Thereafter, a meta-analysis of the search results is performed, showing the trend of publications for each approach. The state of the art of the modelling approaches followed with the evaluation of each approach for future directions is given. At the end in chapter \ref{chap:Conclusion} a brief conclusion is provided.

\label{chap:Review}
\section{Comparison of Cerebral and Abdominal Aortic Aneurysms}
\label{sec:ComparisonAAA-CA}

\begin{figure}[ht]
	\centering
		\includegraphics[width = 0.8\textwidth]{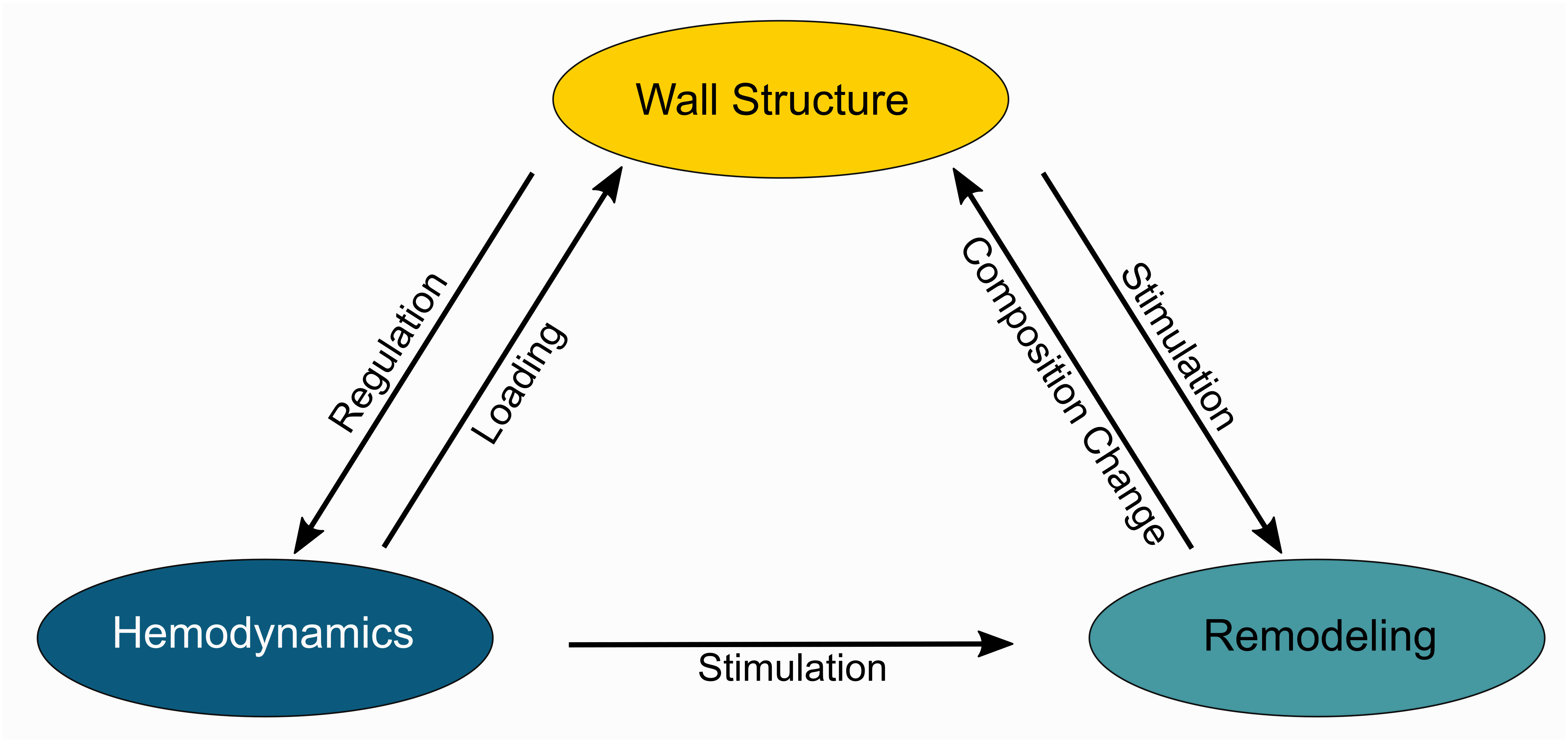} 
	\caption[Influential interaction between wall structure, hemodynamics and remodelling]{Interaction between wall structure, hemodynamics and remodelling process: One of the functions of the vessel wall is regulating blood flow. Hemodynamics in term acts upon the wall structure as stress. Ongoing aberrant hemodynamics and stress sensitive cells inside the wall initiate a remodelling process. As a result of this remodelling, cell composition and therefore wall properties change, ideally to increase wall strength.}
	\label{fig:AneurysmProcesses}
\end{figure}

This section reviews current theories and studies on the formation and development of saccular CAs and fusiform AAAs. After a brief paragraph on the structure of arterial vessels, the mechanisms for aneurysm initiation are discussed. The second part deals with aneurysm growth and development towards rupture. In the third part, the theories on both aneurysm types are brought together to point out similarities and differences. Figure \ref{fig:AneurysmProcesses} depicts the general structures and processes involved in aneurysm initiation and development. Table \ref{Tab:Vergleich} summarises major aspects of abdominal aortic and cerebral aneurysms.

\subsection{Aneurysm Formation}
\label{subsec:AneurysmFormation}

The arterial wall consists of three layers: The tunica intima is the innermost layer, consisting of the endothelial layer (EL) adjacent to the lumen and the internal elastic lamina (IEL) separating it from the tunica media. The tunica media consists of smooth muscle cells (SMC), elastic tissue and collagen, which form the extracellular matrix (ECM). The outermost layer is the tunica adventitia. Larger arterial vessels feature a second external elastic lamina (EEL) between the media and adventitia \cite{Diagbouga.2018}.

\textbf{Formation of Cerebral Aneurysms}

Cerebral vessels have less elastic fibers and a higher amount of elastic lamina compared to the aortic artery while also lacking the EEL \cite{Wilkinson.1972}. Since elastin plays a major role in load baring for the arterial wall \cite{Choke.2005}, this makes cerebral arteries prone to hemodynamic loads and aneurysm formation \cite{Etminan.2014}.

At present, it appears as if the degeneration of the IEL represents a key process prior to the dilation of the arterial wall and subsequent, formation of an aneurysm \cite{Zhang.2019}. The most eminent theory for CA formation emphasizes the significance of acquired lesions \cite{Bacigaluppi.2014} and promotion by genetic deficits of collagen production \cite{Zhang.2014}. \\ 
The initiation of CAs also strongly correlates with hemodynamic factors \cite{Meng.2014}. Previous studies demonstrate that aneurysm initiation is triggered by high wall shear stress (WSS) and a positive gradient of WSS exceeding a certain threshold \cite{Metaxa.2010}. WSS is a drag force exerted by the blood flow upon the vessel wall \cite{Diagbouga.2018}. Endothelial cells (EC), which are present in the intima, are able to sense and respond to hemodynamic forces \cite{Hsiai.2008}. Through endothelial cell mechanotransduction high WSS and a positive wall shear stress gradient (WSSG) lead to proteolytic activity resulting in ECM degeneration and damage of the IEL \cite{Meng.2014}. Hence CA formation typically occurs at bifurcations, where complex flow patterns like flow separation, re-circulation and spatial variations are present \cite{Atlas.1997}. \\

In addition to these aneurysm-specific causes, the occurrence of cerebral aneurysms correlates with female gender, alcohol or nicotine consumption and hypertension \cite{jung2018new}.

\textbf{Formation of Abdominal Aorta Aneurysms}

AAA primarly develop in the most distal abdominal part of the aorta, between the renal arteries and the aortic bifurcation \cite{Lasheras.2007}.  
Compared to cerebral arteries, the aorta has a pronounced elasticity due to a higher ratio of elastin to collagen. This ensures a nearly constant blood flow over the cardiac cycle, which is accomplished by outward movement of the aortic wall during systole and inward realignment during the diastole, the so-called Windkessel effect \cite{Hussein.2017}. Regarding the wall structure of the aorta, the elastin to collagen ratio decreases downstream, which in turn stiffens the wall and compromise its motion \cite{Halloran.1995}. This phenomenon combined with the reflection of pressure pulse waves at the aortic bifurcation makes the abdominal region more susceptible for aneurysm formation as compared to the thoracic aorta \cite{Moore.1992, Dua.2010}. 
Similar to CAs, hemodynamic forces appear to play a key role in the formation of AAAs \cite{Humphrey.2012}. Regions of elevated WSS in the abdominal aorta correlate with regions of dilation and aneurysm formation \cite{Sughimoto.2016}. \\
AAA display a strong correlation with smoking, ageing and sex \cite{Howard.2015,Choke.2012,Humphrey.2008}. It is hypothesised \cite{Lasheras.2007} that the phenomenon stems from the age induced structural changes of the aortic wall. When arteries age, they become stiffer and increase in size. This process mostly takes place in the media, which starts to thin out and lose the orderly structure of the elastic fibers, leading to degeneration of elastin and an increase in collagen \cite{Lasheras.2007}. 

Besides ageing, atherosclerosis is strongly affiliated with the initiation of AAAs \cite{Reed.1992,Kaschina.2009}. Atherosclerosis leads to the formation of plaque upon the vessel wall \cite{Xu.2001}. Rupture of this plaque may cause an inflammatory response, reduction of wall thickness and an increase in wall stress, promoting aneurysm dilatation \cite{Xu.2001}. Recently the causality between atherosclerosis and AAA formation has been challenged by newer studies, claiming rather a parallel development \cite{Johnsen.2010}. Nontheless, the development of AAAs is clearly associated with the alteration of the arterial wall, composition and a degeneration of elastic fibres \cite{He.1994} in conjunction with inflammatory processes which play an important role for the degeneration of tissue \cite{Alexander.2004}. 

\subsection{Aneurysm Growth and Development towards Rupture}
\label{subsec:AneurysmGrowth}

After the initial arterial dilation, most aneurysms change in size and shape through growth, remodelling and distension because of hemodynamic pressure \cite{Frosen.2012}. Growing CAs are 30 times more likely to rupture \cite{Brinjikji.2016} and higher growth rates for AAAs are associated with increased rupture risk \cite{Thompson.2010}. 
Just as in a healthy arterial wall, there exists a balance between mechanisms for repair and proliferation, and processes that degrade and destruct cell structures \cite{Frosen.2012, Humphrey.2012}. An imbalance in favor of cell degradation leads to a weakened and rupture-prone aneurysm wall. \par

\textbf{Development of Cerebral Aneurysms}

For CAs, the newly formed bulge in conjunction, with the loss of the IEL and the hemodynamic pressure lead to an increase in circumferential stress in the aneurysm wall and subsequent distension \cite{Frosen.2019}. To adapt to these circumstances and to increase the wall's ability to withstand stress, SMCs proliferate and synthesize new collagen \cite{Frosen.2004}. Cell proliferation and synthesis of new extracellular matrix is induced by various growth factors produced by macrophages \cite{Frosen.2006}. Macrophages are inflammatory cells, which predominantly secrete matrix metalloproteinase (MMP) of type 2 (MMP-2) and 9 (MMP-9) into the aneurysm wall \cite{Aoki.2007}. As part of the family of proteinases, they degenerate cells of the ECM, notably collagen and elastin \cite{Murphy.1995}. Therefore, the presence of macrophages in high numbers and their influence on cell proliferation facilitates growth of the aneurysm wall, while this presence at the same time weakens the wall due to the secretion of MMPs \cite{Frosen.2019}.

Additionally, it activates a feedback loop, which further amplifies macrophage activity and increases expression of monocyte chemoattractant protein-1 (MCP-1) \cite{Aoki.2011}. MCP-1 is a kind of cytokine, that recruits monocytes, which differentiate into macrophages, to the place of inflammation in the aneurysm wall. It was demonstrated, that blockage of MCP-1 in rats resulted in lower macrophage concentration and decreased aneurysm growth \cite{Aoki.2009}. Ongoing growth of the CA and the subsequent increase in diameter lead to a higher circumferential wall stress \cite{Frosen.2019}. However, some CAs have regions in which SMCs are partially absent \cite{Frosen.2012} or have impaired function \cite{Ollikainen.2016}. This inhibited ability to adapt to increased wall stress might cause an imbalance towards cell degradation and render these focal wall regions more prone to aneurysm rupture \cite{Frosen.2019}. \par 

It is assumed that hemodynamics plays an important role during aneurysm growth \cite{Cebral.2017, Longo.2017, Castro.2013b}. The issue whether high or low WSS is the main hemodynamic factor contributing to aneurysm growth \cite{AcevedoBolton.2006, Sugiyama.2012} and rupture \cite{Cebral.2011, Xiang.2011, Castro.2009} is cause for an ongoing debate. As a unifying solution, Meng et al. \cite{Meng.2014} proposed that both conditions drive different pathways that lead to degradation of the ECM and facilitate aneurysm rupture. On one hand, high WSS and a positive WSSG can occur near the aneurysm dome when an impinging flow carrying high inertia enters the sac. It is likely that similar inflammatory remodelling mechanisms leading to aneurysm formation are relevant for aneurysm growth, too. They associate this driveway with a phenotype of aneurysms small in size and a mostly thin, translucent wall \cite{Meng.2014}. Additionally, areas of high WSS are associated with the formation of blebs, which are secondary, smaller outpouches in the aneurysm wall \cite{Cebral.2010}. On the other hand, an increase in aneurysm size and recirculating flow inside the aneurysm lead to low and oscillating WSS. Under these conditions, the EL responds with increased permeability to the media layer, up-regulation of surface adhesion molecules and cytokines in the aneurysm wall \cite{Chiu.2011}. These flow conditions also lead to an increased residence time of blood which causes elevated transmigration of leukocytes into the wall \cite{Meng.2014} and facilitate the formation of atherosclerotic plaques \cite{Frosen.2012}. As a result, proteolytic activity and ECM degradation is increased leading to growth and a weakened aneurysm wall. Additionally, slow and oscillating flow and the subsequent higher blood residence time predispose to intraluminal thrombus (ILT) formation. Erythrocytes, leukocytes and platelets get trapped inside the thrombus and release growth factors and cytokines. Together with inflammatory thrombus cells these infiltrates promote synthesis of SMCs and secretion of MMPs, which drive collagen degradation \cite{Frosen.2012}. The above mentioned pathway is associated with a phenotype of large CAs with thick walls interspersed with atherosclerotic plaques \cite{Meng.2014}. These phenotypes of CAs represent the upper and lower end of a spectrum. Resulting from the interplay of pathways with high and low WSS \cite{Meng.2014}, most CAs are in between both extremes and show mixed characteristics \cite{Kadasi.2013}. \par

\textbf{Development of Abdominal Aortic Aneurysms}

While the development of AAA is characterised by a loss of the elastic lamina and SMCs, its enlargement is attributed to a higher collagen turnover \cite{Humphrey.2012}. With the elastin typically tasked with load bearing, its loss has to be compensated by additional collagen synthesis \cite{Choke.2005}. This is supported by studies demonstrating the increase in collagen content with larger AAA size \cite{Minion.1994, Choke.2005}. Additionally, the remodeling process leads to differences in collagen cross-linking, alignment and distribution as compared to the healthy aorta. Both processes subsequently influence the distensibility of the wall \cite{Humphrey.2012}. 

Inflammation induced by a immune system response plays a key role in AAA progression. Leukocytes, particulary neutrophils and macrophages, are observed in the tunica media and adventitia of AAA \cite{Maguire.2019}. Leukocytes enable pathways into the AAA wall for additional inflammation cell, which produce proteolytic enzymes like MMPs. Just as in CAs, MMP-2 and MMP-9 are responsible for SMC degradation and subsequent wall weakening. Reactive oxygen species (ROS) are an additional mechanism shown to facilitate SMC degradation and ECM remodelling \cite{Maguire.2019}. Normally MMP activity is regulated by tissue inhibitors of MMPs (TIMP), which are found in a lower ratio with MMP in AAAs compared to the normal aorta \cite{Tamarina.1997}. While MMPs can also drive ECM synthesis, upregulated MMP production and high MMP/TIMP ratio point towards an imbalance towards ongoing ECM degradation. More about the role of enzymes in AAA development can be found in \cite{Maguire.2019}.

In line with CAs, hemodynamics plays an important role in AAA development. The initial dilation in the abdominal aorta influences the blood flow through the abdominal aorta. A widened aorta leads to adverse pressure gradients, flow separation from the wall, formation of recirculation zones and vortex formation \cite{Bauer.2020}. As a result, low WSS and complex flow are predominant in AAAs, with possible transition from laminar to turbulent flow \cite{Bauer.2020}. In contrast, high WSS values can be observed at the proximal and distal neck of the AAA, due to vortex formation and impingement. Consequently, these areas undergo high temporal changes in WSS \cite{Bauer.2020}. Low, oscillating WSS may correlate to increased leukocyte density and infiltration into the AAA wall \cite{Dua.2010}. Additionally, recirculating flow conditions increase the residence time of blood near the wall, which facilitates ILT formation and growth \cite{Basciano.2011}. In contrast to CAs, ILTs are present in nearly all large AAAs \cite{BehrRasmussen.2014}. The formation and development of an ILT is agreed to have a significant influence on AAA development. On a mechanobiological level, the ILT is an additional source of proteinase, which may lead to increased ECM degradation \cite{Tong.2015}. The wall covered by ILT is reported to be thinner with less elastin, shows more signs of inflammation \cite{Kazi.2005} and suffers from local hypoxia in regions with thicker ILTs \cite{Vorp.2001}. Therefore, ILT presence and development may favor formation of structural instabilities in the AAA wall and lead towards rupture \cite{Tong.2015}.\par

\textbf{On Rupture, Similarities and Differences between CAs and AAAs}

It is assumed that rupture represents a local phenomenon, occurring at sites, where wall stress exceeds the wall strength \cite{Vorp.2007}. It appears that for CAs and AAAs the mechanisms mentioned above, especially proteolytic cell degradation, weaken the aneurysm wall over time, until it is unable to bear the hemodynamic load. More on the biomechanical factors that seem to correlate with aneurysm rupture is found in Sec.~\ref{sec:SystematicModelReview}.

\begin{table}[htb]
	\centering
	\fontfamily{bch}
	\begin{tabular*}{0.9\textwidth}{@{}l@{\extracolsep\fill}p{4.7cm}l@{\extracolsep\fill}p{7cm}@{\extracolsep\fill}}

\textbf{} &	\textbf{Abdominal Aortic} &	\textbf{Cerebral}\\
Predominant shape & fusiform \cite{Humphrey.2008} & saccular \cite{Humphrey.2008} \\[0.1cm]
Share of aneurysm deaths &  {40}{\%} \cite{WHO_Database.2020} & $\approx{38.7}{\%}$ \cite{WHO_Database.2020} \\[0.1cm]
Primarily affected sex & men \cite{WHO_Database.2020} & women \cite{WHO_Database.2020}\\[0.1cm]
Genetic risk factors &  \multicolumn{2}{l}{Ehlers-Danlos and Marfan syndrome \cite{Hoskins.2017}}  \\[0.1cm]
Environmental risk factors & \multicolumn{2}{l}{smoking, alcohol consumption, hypertension \cite{Lasheras.2007}}  \\[0.1cm]
&&\\
Vessel structure & more elastic fibres \cite{Wilkinson.1972} & less elastic fibres \cite{Wilkinson.1972} \\ 
                          &  presence of EEL \cite{Diagbouga.2018}  & lack of EEL \cite{Diagbouga.2018} \\
Growth factors & complex hemodynamics & high and low WSS trigger \\
&	vortex formation  &   	separate pathways\\
&	inflammation and activation  &   of increased MMP  \\ 
&	of MMPs, collagen turnover	&	production \cite{Meng.2014} \\ 
&	\cite{Bauer.2020, Tanweer.2014, Maguire.2019, Humphrey.2012} &	 \\ [0.1cm]
&&\\
Responsible clinicians & vascular surgeons & neurosurgeons, radiologists \\[0.1cm]
Clinical Imaging & ultrasound, CT, MRI \cite{Moll.2011} & CT, DSA, MRI \cite{Brinjikji.2018} \\[0.1cm]  
Critical diameter & $ {>50}${mm} \cite{Moll.2011} &	 ${>{12}}${mm} \cite{Williams.2013}\\[0.1cm] 
Rupture risk [p.a.] &  {3}-{9}{\%} \cite{Chalouhi.2013, Humphrey.2008} & $<{1}${\%} \cite{Wiebers.2003}\\[0.1cm] 
Mortality after rupture & {65}-{80}{\%} \cite{Humphrey.2008} & {45}{\%} \cite{Hoskins.2017} \\[0.1cm] 

	\end{tabular*}
	\caption{Comparison of characteristics between AAA and CA. Share of aneurysm deaths and primarily affected sex is related to data on deaths caused by aneurysms in Germany between 2006 and 2015.}
	\label{Tab:Vergleich}
\end{table}

Two of the major differences between CAs and AAAs are the dominant geometry, saccular versus fusiform, and the size of the host artery in which they form. Both implicate different flow characteristics influencing the aneurysm. However, hemodynamics and especially WSS play a key role in formation and development of both aneurysm types. For CAs, high WSS near bifurcations in the circle of Willis facilitate aneurysm formation, while for AAAs low WSS in combination with flow recirculation is more significant. In growing aneurysms, both CAs and AAAs, areas of low and high WSS are present. These may drive different pathways of remodeling mechanisms, leading towards a weakened aneurysm wall. 

Similarities can be found in the remodelling processes of the aneurysm wall. Inflammation is a key mechanism during the development of both aneurysm types and at least for the formation of AAAs. Similar observations of early loss of load bearing elastin and subsequent synthesis of collagen to stabilise the aneurysm wall have been made. Especially the role of MMPs as main source of proteolytic activity and subsequent ECM degradation is highly relevant in CAs and AAAs. Additional sources of wall weakening enzymes like ILTs in AAAs or atherosclerotic plaques found in some CAs, might further imbalance local cell proliferation and degeneration, creating rupture-prone hot spots in the aneurysm wall.

Table \ref{Tab:Vergleich} contrasts the major characteristics of AAA compared to CA.

\section{Systematic Model Review}
\label{sec:SystematicModelReview}
This section includes a systematic model review on the most relevant modelling approaches for aneurysms. Section \ref{subsec:DescripModellApproach} gives a general description of the considered approaches, while in section \ref{subsec:SearchStrategy} the  search strategy applied is explained. Resulting from this search a Meta analysis was performed described in \ref{subsec:MetaAnalysis} and current state of the art models for each modelling approach were identified further explained in section \ref{subsec:StateOfTheArt}.

\subsection{Description of Considered Modelling Approaches}
\label{subsec:DescripModellApproach}

Assessing literature for aneurysm rupture risk evaluation and aneurysm modelling, the four most prominent modelling approaches can be identified as: computational fluid dynamics (CFD), finite element analysis (FEA), assessment-tools \& dimensionless parameters (AT\&DP) and machine learning (ML). Additionally, fluid-structure interaction (FSI) and fluid-solid-growth (FSG) models are frequently discussed in aneurysm research, which justifies a closer look on these approaches as well. This section explains their basic principles of the different approaches to facilitate a common understanding and help to understand current state of the art models presented in \ref{subsec:StateOfTheArt}.

\begin{enumerate}
\item \textbf{Finite Element Analysis (FEA)}

Finite element analysis describes the numerical simulation of phenomena related to structural mechanics based on the finite element method (FEM). It is widely used in various engineering disciplines to evaluate the reaction of materials and structures to external and internal forces. For the example of aneurysms, images received from CT or MRI scans are computationally analysed during the segmentation process to distinguish between different anatomical structures like arterial wall, ILT and surrounding tissue. Next, a digital reconstruction of the aneurysm, called mesh, is generated. Instead of a continuous structure, the real aneurysm is represented by a multitude of discretised finite elements. The physical behaviour of these elements is described via partial differential equations, that take into account external forces and the behaviour of surrounding elements. These constitutive equations model the behaviour of the real structure and consider for characteristics like elasticity and isotropy. Combination of the response of each element to external forces gives information about the whole structure regarding stress, strain or deformation. A description of this method and explanation for relevant terms can be found is given by Gasser \cite{Gasser.2016b}.

\item \textbf{Computational Fluid Dynamics (CFD)}

Computational fluid dynamics refers to software tools solving the Navier-Stokes equation numerically. The Navier-Stokes equation is a set of differential equations derived from Newton's second law of motion, which states that the rate of change of momentum of a body is directly proportional to the force applied. The Navier-Stokes equation therefore describes the motion of viscous fluid particles. For most problems, the Navier-Stokes equation cannot be solved analytically. Consequently CFD-solvers, providing numerical solutions of the equation, have risen to prominence in the last few decades. Besides simulating the flow field, CFD-software is used to compute interactions between fluids and the surrounding structures using proper boundary conditions. In context of aneurysm research, CFD-software is mainly used to compute the hemodynamic forces and the flow field of blood during a cardiac cycle. For a detailed description of an CFD model employed for CAs and its governing equations see Campo-Deaño et al. \cite{CampoDeano.2015}.

 \begin{enumerate}
 \item \textbf{Fluid-Structure Interaction (FSI)} 
 
 Fluid-structure interaction describes a type of modelling approach that uses a combination of FEA and CFD software tools. Here, FEA gives information on the deformation of the arterial wall, while CFD simulates the blood flow through the lumen. Because wall deformation influences the hemodynamic forces and vice versa, the FSI based approach considers both with the intent to create a more realistic model. Teyduzar et al. \cite{Tezduyar.2011b} gives a deep insight into FSI models, exemplarily for CAs. 
 
 \item \textbf{Fluid-Solid-Growth (FSG)}
 
 Fluid-solid-growth models combine the FSI approach and the mechanobiological approach of growth and remodelling (G\&R). Hence, they consider the hemodynamic forces, wall stress and deformation and remodelling processes that capture long time changes in the wall structure due to mechanobiological mechanisms. FSG can be considered the most realistic and complex model approach, as these three processes are presumed to be the most significant regarding influence on aneurysm development and rupture. Of special interest is the coupling between the short-term FSI model, which simulates the aneurysm over a cardiac cycle with the long-term simulation of G\&R processes. The term FSG was first used by Humphrey and Taylor \cite{Humphrey.2008} and their paper provides an overview over the processes involved.
 \end{enumerate}
 
\item \textbf{Assessment-Tools \& Dimensionless Parameters (AT\&DP)}

AT\&DP consists of assessment-tools and dimensionless parameters. Dimensionless parameters are derived from complex models and are used to detect aneurysms and aneurysm rupture \cite{Asgharzadeh.2019}. They are calculated by analysing medical images or by processing results from CFD, FEA, FSI or FSG simulations. 

Assessment tools have been designed to help medical personnel in speeding up their decision making. They consider a set of factors which corresponds to a particular rupture risk rate. These were defined based on extensive research and experimentation. They are only recommended for use if they prove a higher prediction quality than the diameter criterion. Assessment-tools act as questionnaires based on statistical information that conclusively predict the likelihood of an aneurysm rupture. They require patient’s data such as age, gender, ethnicity, etc. while the information regarding aneurysm size is extracted with the help of medical imaging techniques. Assessment-tools either predict the rupture risk or advise a particular treatment which in turn is directly dependent on the rupture risk \cite{Etminan.2014}. These tools use data that is readily available in hospitals and from the patient’s history. 

\item \textbf{Machine Learning (ML)}

Artificial intelligence (AI) is “the science and engineering of making intelligent machines, especially intelligent computer programs” \cite{.2004}. Machine learning is a sub-discipline of AI that combines computer science with statistics that is “concerned with the question of how to construct computer programs that automatically improve with experience” \cite{Mitchell.2010}. ML enables the computer to recognise statistical patterns within data. In comparison to traditional programming, ML does not deliver the output for given data and a given function, but it delivers the function that can be used to predict these outputs for future data. 
In general, the performance of ML algorithms improves with experience. This means the more data the algorithm receives the better it becomes and the more accurate it can be in its predictions. Classical ML algorithms such as Random Forests and Support Vector Machines, are fed with structured data. To apply ML directly on image data, usually deep learning is used. Deep learning is a sub-discipline of machine learning that uses artificial neural networks (ANN). 

In the context of aneurysms, ML is based on structured patient data, clinical images or on a combination of both. Here, ML is mostly used for image segmentation, aneurysm growth prediction or to directly estimate the rupture risk related to a specific aneurysm.

\end{enumerate}

\subsection{Search Strategy}
\label{subsec:SearchStrategy}

The search strategy to identify scientific records to be considered for this review was implemented on the Web of Science - Database while the systematic paper review was conducted on the basis of the guidelines as described by Preferred Reporting Items for Systematic Reviews and Meta-Analyses (PRISMA). 

The library used for the literature search was Web of Science, because it incorporates an extensive amount of databases compared to other libraries and allows to set up user-specific search queries. Furthermore, Web of Science allows clustering papers based on keywords, relevance or citations and it presents a graphical representation of the publications per years \cite{Falagas.2008}. The initial search was conducted using the Web of Science library on June 20$^{th}$, 2020 with the query shown in the appendix in table \ref{Tab:Datenbankabfrage1}.

The query is designed in a way to include as many research papers as possible that deal with aneurysm modelling and simulation. Based on the importance of CFD, FEA, AT\&DP and ML for aneurysm modelling, keywords related to these approaches were explicitly included in the search query. Because aneurysm computational modelling, especially with regards to rupture risk prediction is the focus of this review, the query sorts out papers dealing with experimental aneurysms in animals and pre- or post-aneurysm rupture treatment solutions.

\begin{figure}[h]
	\centering	
		\includegraphics[width=0.8\textwidth]{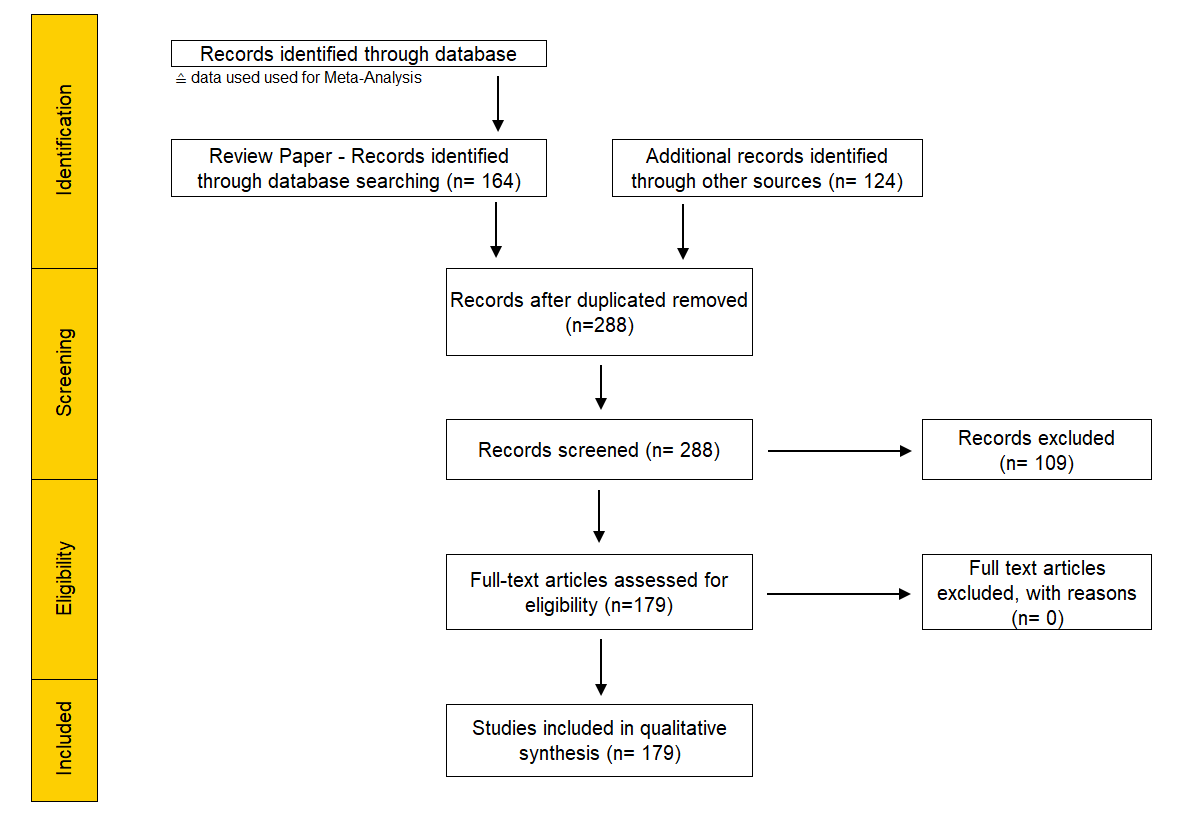}
	\centering	
	    \caption{PRISMA - Flow Diagram}
	\label{fig:PRISMA_FLOW_DIAGRAM}	
\end{figure}

The query-based Web of Science search delivered 3127 papers. These 3127 papers were further clustered to perform a meta-analysis described in \ref{subsec:MetaAnalysis}. Restricting the Web of Science search on review papers and full English publications only, 164 records were left. These 164 review papers formed the basis for this review paper and were screened based on publication titles and their corresponding abstracts by the researchers (AZ, JB, MIAK, and ST). Research papers vital for the research topic that were not found using the above-mentioned query were identified and consequently appended to our list as additional records identified through other sources. Once a paper had fulfilled the inclusion criteria, the full text was examined. Figure \ref{fig:PRISMA_FLOW_DIAGRAM} depicts the PRISMA flow diagram.

\subsection{Meta-Analysis}
\label{subsec:MetaAnalysis}

Having identified publications to consider within this review by a systematic database search and having identified four major modelling approaches to cluster for, a meta-analysis is performed to gain more information about the scientific relevance of each approach. To identify the modelling approaches of major relevance and the preferred approaches for abdominal and cerebral aneurysms respectively, each of the 3127 records obtained from the database search is further clustered by modelling approach and aneurysm location when applicable. Abdominal aortic and thoracic aortic aneurysms were counted for abdominal aneurysms in this context. Further information about the conduction of the meta-analysis can be found in appendix \ref{S1_Fig}. Figure \ref{fig:PublicationLocation} shows the relevance of the modelling approaches for abdominal and cerebral aneurysms. 

\begin{figure}[h]
	\centering
		\includegraphics[width=1\textwidth]{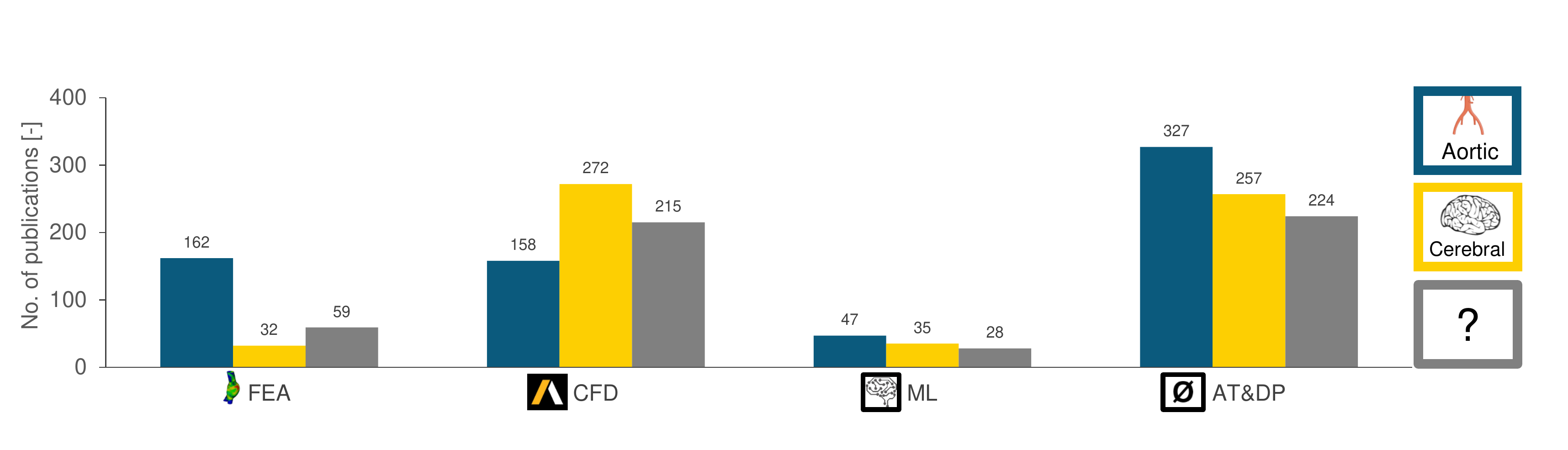}
	    \caption{Number of publications found during the systematic review, clustered by modelling approach and aneurysm location. \\Blue: aortic aneurysms, yellow: cerebral aneurysms, grey: unspecified aneurysm location in title of publication}
	\label{fig:PublicationLocation}
\end{figure}

The meta-analysis shows that CFD modelling plays a major role for modelling of cerebral aneurysms, whereas FEA is primarily relevant for aortic aneurysms. This correlation is again observed during review of the content of aneurysm literature. Due to hemodynamic peculiarities in CAs, CFD modelling plays a key role there, whereas in abdominal aneurysms hemodynamics is of secondary importance compared to the structural wall behaviour, depicted by FEA. This trend of modelling cerebral aneurysms with CFD approaches, whereas FEA is applied for aortic aneurysms is depicted by state of the art models for rupture risk evaluation (\ref{Tab:ModelOverview}), too. 
ML and AT\&DP approaches are of almost equal importance for cerebral and aortic aneurysms. With a total of 808 publications clustered to AT\&DP this category is certainly important, but the number of records clustered for this category must be considered carefully, because parameter studies performed for CFD or FEA models are likely to be included in this category as well. This can be understood regarding the design of the meta-analysis described in appendix \ref{S1_Fig}. Overall, there is about an equal amount of publications for aortic and cerebral aneurysms and CFD related modelling approaches play a major role in aneurysm modelling. 

\begin{figure}[h]
	\centering
		\includegraphics[width=1\textwidth]{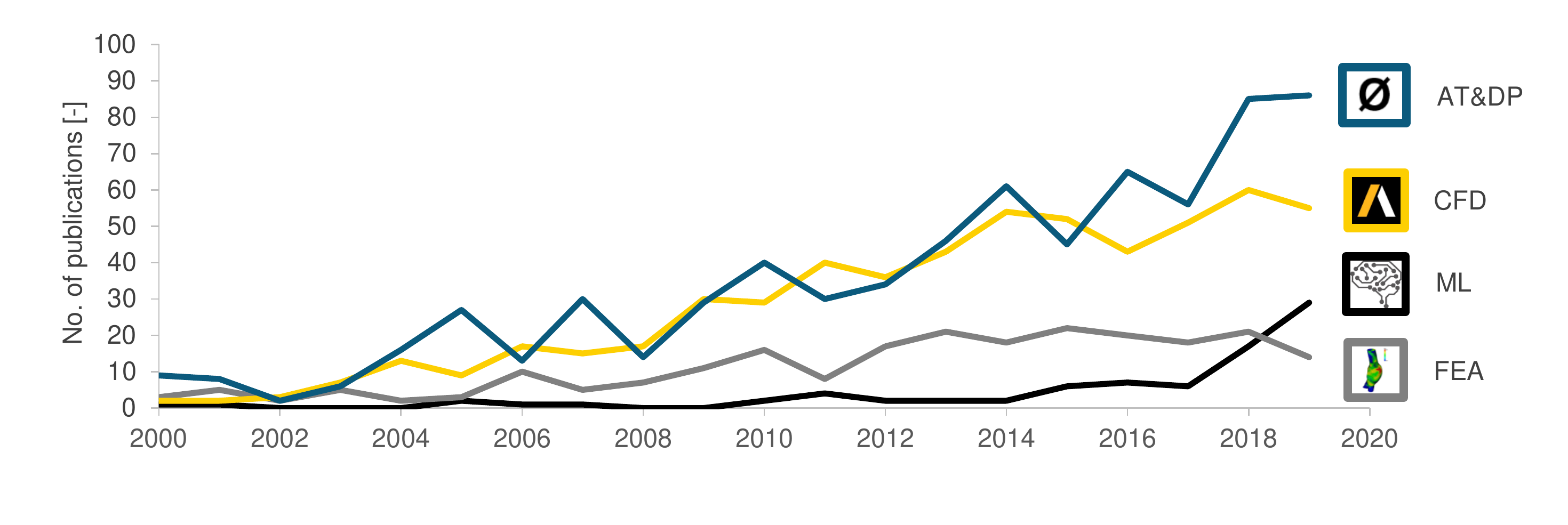} 
	\caption{Development of the number of publications between 2000 and June 2020. Publications are clustered by modelling approach.}
	\label{fig:PublicationHistory}
\end{figure}

To determine which modelling approaches have potential for clinical aneurysm risk evaluation, analysing current scientific aneurysm modelling trends is crucial. Figure \ref{fig:PublicationHistory} depicts the result of this analysis. It is observed that FEA models have passed their peak and show a declining trend in annual publications. The number of publications for ML increased significantly during the past years. With 25 publications for this model class already by June 20$^{th}$, 2020 this trend is likely to continue. AT\&DP have a constantly high significant scientific relevance based on the number of publications during the past years. As mentioned, the AT\&DP relevance in this analysis has been most likely overestimated. Therefore, it can be concluded that CFD models are currently the major scientific focus for aneurysm modelling, but ML based models are becoming increasingly relevant. Nevertheless, figure \ref{fig:PublicationHistory} is not only based on records dealing with aneurysm rupture risk evaluation but includes publications for other types of aneurysm modelling, like for example research on aneurysm growth, too. Therefore, it cannot be directly concluded that CFD models play the major modelling role for aneurysm rupture risk evaluation. In general, CFD related approaches are frequently applied methods to gain further understanding in aneurysm growth mechanisms for both, aortic and cerebral aneurysms, because of the importance of hemodynamics in the context of growth. 

\subsection{State of the Art of Aneurysm Modelling Approaches}
\label{subsec:StateOfTheArt}
\begin{figure}[h]
	\centering
		\includegraphics[width = 0.8\textwidth]{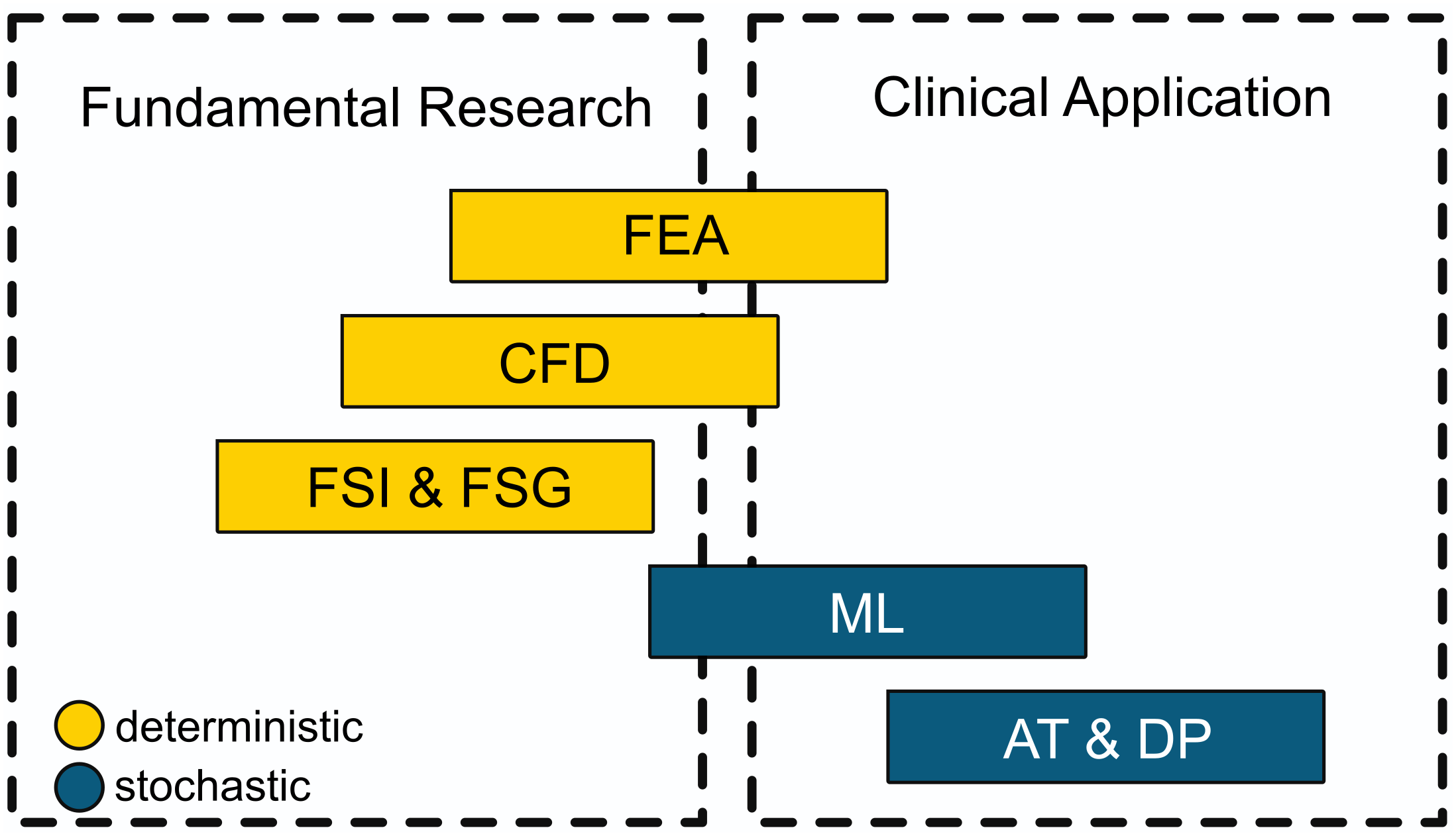} 
	\caption[Categorisation of relevant model categories]{Categorisation of deterministic (yellow) and stochastic (blue) modelling approaches: While the four different deterministic approaches are more often used in fundamental research, stochastic models are primarily for clinical application.}
	\label{fig:ModelCategorisation}
\end{figure}

Based on the systematic literature search described in \ref{subsec:SearchStrategy}, this section summarises the latest state of the art of the four major modelling approaches (FEA, CFD, AT\&DP, ML) and the combined approaches (FSI and FSG). The evaluation of each category focuses on the research regarding aneurysm rupture and potential implementation for rupture risk prediction in a clinical environment. Deterministic approaches, namely FEA, CFD, FSI and FSG, research on processes and biomechanical factors that are thought to influence aneurysm rupture. Information on these biomechanical risk factors is included in this review, because they are relevant for a better understanding of aneurysm rupture. On the other hand, ML and AT\&DP as stochastic approaches are mostly aimed at clinical implementation and do not consider underlying processes in aneurysm development. As a result, this review contains more content to deterministic approaches compared to the stochastic ones, which should not express a difference in their importance. Above mentioned modelling approaches can be categorised due to their priority application in fundamental research or clinics, as seen in figure \ref{fig:ModelCategorisation}.

\subsubsection{Finite Element Analysis}
\label{subsubsec:FEAstate} 
Finite element analysis is used to research on biomechanical factors that influence the development and rupture of aneurysms. The focus of this approach is to predict the stress exerted on the aneurysm wall and the wall's ability to withstand it, the so called wall strength. These characteristics, wall stress and strength, although often defined differently by different authors, can be used to make assumptions about the rupture risk and sometimes about its probable location. Stand-alone FEA, without combination of G\&R or CFD based modelling approaches, is used mainly for AAAs and the there predominant fusiform geometry (compare to section \ref{sec:ComparisonAAA-CA}). Therefore, reviews and additional papers used to evaluate the state of the art of FEA deal exclusively with fusiform AAAs. \par 

While for specific purposes generic aneurysm models are adequate, for research on biomechanical factors and rupture risk prediction most FEA models aim to be patient-specific. This raises the challenge to find a compromise between realistic model assumptions and a reasonable computational effort. A patient-specific FEA model requires information about the individual AAA geometry, characteristics of the wall tissue and hemodynamic load, as well as a appropriate relation between wall stress and deformation \cite{Malkawi.2010}. Geometric features and a precise AAA reconstruction is considered to have the largest influence on an accurate model prediction \cite{Raut.2013}. Current approaches emphasise the need for inclusion of the ILT, due to its high occurrence and recognised influence on growth and rupture mechanisms \cite{Raut.2013}. \par

Before a finite element analysis can take place, the realistic AAA geometry has to be obtained and reconstructed to a digital model, called mesh. The most common method for AAA imaging is computerised tomography (CT). Magnetic resonance imaging (MRI) and three dimensional ultrasound gain increased importance because of high cost and potential health risks of repeated CT scans \cite{Salman.2019}. From these images of the patient's aneurysm, the digital model is created with segmentation and optimisation software. Exemplary software products for these tasks can be found in Salman et al. \cite{Salman.2019} and recommendations for geometry reconstruction in Raut et al. \cite{Raut.2013}. Based on the geometric reconstruction and model equations for wall stress and strain relationship, most FEA computations calculate the peak wall stress (PWS) to give an indication about rupture risk. PWS is the distribution of circumferential stress over the aneurysm wall, which occurs during maximum systolic pressure \cite{Georgakarakos.2011}. It was shown that PWS has a higher sensitivity (true positive rate) and specificity (true negative rate) compared to AAA diameter when predicting risk of rupture \cite{Fillinger.2003}. This intensified research into wall stress distribution and its potential as additional or alternative clinical indicator for rupture risk. In comparison, WSS, which is highly relevant in AAA initiation and growth as well as in CFD analysis for CAs (see section \ref{subsubsec:CFDstate}), is considered less relevant for AAA rupture prediction. Supportive arguments are the order of magnitude higher circumferential wall stresses compared to the shear stresses and the characteristics of final stage AAAs, namely missing shear sensitive endothelial layer and the ILT, which forms a buffer zone between blood flow and aneurysm wall \cite{Vorp.2007}. \par

The following paragraph summarises relevant biomechanical factors with presumed influence on wall stress or strength. Moreover, assumptions and simplifications are described, when factors are used in FEA modelling:\par

\textbf{Biomechanical factors related to aneurysm geometry}

Besides AAA size, described by its maximum diameter, the individual aneurysm shape is considered to have a great influence on stress and strength distribution of the wall. AAA centerline asymmetry shows a positive correlation with wall stress distribution along the posterior aneurysm wall \cite{Doyle.2009} where the majority of AAA ruptures occur \cite{Darling.1977}. Mean and maximum centerline curvature display a significant correlation with PWS values \cite{Giannoglou.2006}, so does tortuosity, expressed as the increase in length of the vessel compared to a straight line \cite{Georgakarakos.2010}. Raut et al. \cite{Raut.2013} explain a possible correlation of wall areas of high wall stress and local surface saddle points. As to branching arteries of the abdominal aorta, an increase in the iliac bifurcation angle correlates with higher wall stress and WSS \cite{Xenos.2010, Drewe.2017}. A greater proximal neck angle of the AAA produced increased WSS values, indicating an influence in AAA development \cite{Drewe.2017}. Local wall defects, for example blebs, affect the wall strength and therefore the likely location of rupture, which does not always coincide with the area of the highest wall stress \cite{Georgakarakos.2010b}. These results highlight the importance of considering the patient-specific AAA shape, when wall stress, strength and rupture risk are assessed. Moreover, the in early works often used stress calculation via Law of Laplace, which is only valid for spherical or cylindrical shape, is considered to be outdated \cite{Vorp.2007, McGloughlin.2010, Indrakusuma.2016}. \par

\textbf{Biomechanical factors related to wall characteristics}

Characteristics of the aneurysm wall and their implementation in the AAA model, directly effect the realistic and predicted wall strength. The used constitutive model for the AAA wall, assumptions about homogeneity, isotropic behaviour and wall thickness and inclusion of an ILT have a great impact on the results and the validity of specific modelling approaches. AAA wall thickness is non-uniform and experimental values range from {0.23}{mm} to {4.26}{mm} \cite{Raghavan.2006}. For the normal aorta, the posterior wall is thinner than the anterior, likely due to the perivascular support from the spine \cite{Humphrey.2012}. With the observation of a reduction in wall thickness near the rupture location \cite{Raghavan.2006}, one can speculate, if the thinner posterior wall correlates with its increased rupture occurrence \cite{Darling.1977}. Despite its non-uniformity, many FEA models are based upon uniform wall thickness \cite{Humphrey.2012}. However, the assumed wall thickness value influences PWS calculation \cite{Venkatasubramaniam.2004}. A variable wall thickness can be implemented by measuring local values from CT images and following generation of a approximated distribution \cite{Joldes.2017} Measuring the exact patient-specific wall thickness is currently not possible, due to the lack of appropriate noninvasive imaging technology. Wall compliance describes the vessel's ability to stretch during the cardiac cycle and is represented by wall stiffness or strain \cite{Indrakusuma.2016}. AAAs are associated with higher and with age more drastically increasing stiffness compared to the normal aorta, potentially due to loss of elastin \cite{Vorp.2007, Lanne.1992}. There seems to be no correlation with wall compliance and prediction of AAA rupture \cite{Sonesson.1999, Wilson.2003} and conflicting correlation with growth \cite{Wilson.1999, Hoegh.2009}. Wall compliance is modelled with various constitutive equations relating the external stress from the hemodynamic load to the wall strain. Ex vivo measurements on AAA wall tissue suggest a nonlinear stress-strain relationship \cite{He.1994} and an increase in mechanical anisotropy, here stiffening in the circumferential direction, with progressing wall degradation \cite{VandeGeest.2006}. While non linear behaviour is widely used in FEA, most models apply isotropic, hyperelastic constitutive models instead of anisotropic ones \cite{Humphrey.2012}. Alternative anisotropic models have been proposed \cite{Gasser.2006, ZeinaliDavarani.2009}. Assumed anisotropy yields in higher PWS compared to isotropy \cite{VandeGeest.2008}. The aneurysm wall is a three-layered structure with different mechanical properties in each layer, as described in \ref{sec:ComparisonAAA-CA}. This inhomogeneity is not considered in most FEA models \cite{Humphrey.2012}. Vascular tissue, like the AAA wall, is assumed to be incompressible \cite{Carew.1968}. 
Nowadays the inclusion of the ILT in FEA is widely accepted, when a accurate stress distribution is aimed for \cite{Vorp.2007}. While all ILT are considered to be isotropic and inhomogeneous, mechanical properties differ for different types \cite{OLeary.2014}. Its influence on wall stress and rupture risk prediction is not fully understood and still debated \cite{Raut.2013, Kontopodis.2015b}. In FEA, the ILT is often considered with a homogeneous, incompressible, hyper- or viscoelastic constitutive model \cite{Raut.2013}. Just like for the ILT, calcifications localised in the tunica media are debated about in regards to their impact on stress distribution. Different model approaches and challenges in identifying calcifications during image segmentation complicate a definitive evaluation \cite{Raut.2013}. \par

\textbf{Biomechanical factors related to the hemodynamic load}

Stress on the aneurysm wall can be exerted by three sources: the hemodynamic load acting on the luminal surface, perivascular tissue in contact with the outer wall surface and pre-stretch, which acts in axial direction \cite{Humphrey.2012}. Most FEA models only consider the hemodynamic load, in the simplified form of blood pressure. If available, patient-specific systemic blood pressure during systolic phase is used to calculate wall stress, instead of a local aortic pressure, which cannot be obtained via noninvasive methods \cite{Malkawi.2010}. Other models use the mean arterial pressure (MAP) \cite{Gasser.2010}. The blood flow through the aorta is by nature pulsatile, which translates to a nonuniform pressure function over the cardiac cycle. Addition of the flow to model computation leads to FSI models (see section \ref{subsubsec:FSIstate}). For sole FEA models, use of a nonuniform pressure distribution instead of a static value can improve wall stress calculation without the need for complex FSI computations \cite{Raut.2013}. Caution must be applied in regards to the prestressed condition of aneurysms, which are always at least under diastolic pressure. When the scan images are acquired and a geometric reconstruction is created, it does not show the unloaded state at zero internal pressure, but the prestressed state. Application of the blood pressure as boundary condition for the prestressed AAA model leads to overestimation of the PWS \cite{Raut.2013}. Various numerical methods have been developed to approximate the unloaded aneurysm geometry from prestressed reconstructions \cite{Lu.2007, Putter.2007, Raghavan.2006b}.\par

\textbf{Indicators for rupture risk assessment}

A longstanding goal of the biomechanical research community is the development of parameters and assessment tools for rupture risk prediction and their usage in a clinical environment. Multiple of such parameters are grouped with the term "biomechanical rupture risk assessment" (BRRA) method \cite{Gasser.2016b}. One of the first is the above mentioned PWS, which is also the foundation for stress analysis in later progressed parameters \cite{Fillinger.2002, Fillinger.2003}. To consider the important aspect of wall strength, the local rupture potential index (RPI) is introduced \cite{VandeGeest.2006b}, its maximum also termed "peak wall rupture risk" (PWRR) \cite{Gasser.2010} or "peak wall rupture index" (PWRI) \cite{Gasser.2016b}. It is calculated as the dimensionless ratio between PWS and wall strength. Due to the inability to measure wall strength in vivo, a statistical model is applied, which considers a patient's local ILT thickness, local normalised AAA diameter, family history and gender \cite{VandeGeest.2006b}. PWRI is able to differentiate between diameter matched ruptured and intact AAA with statistical significance \cite{Gasser.2010}. A third index belonging to BRRA is the probabilistic rupture risk index (PRRI) \cite{Polzer.2015}. Instead of the deterministic approach of PWRI it utilises a probabilistic method to account for uncertainties in wall stress and strength predictions as well as the stochastic nature of failure \cite{Polzer.2015}. In a quasi-prospective validation study \cite{Polzer.2020} both PRRI and PWRR, a scaled version of the deterministic PWRI, were tested on their ability to predict AAA rupture within different intervals after a CT scan of patient's intact AAA. The results were compared to assessments with maximum diameter and sex-adjusted diameter, which reflects the higher AAA rupture risk in women. Validation criteria was the area under the curve (AUC) of a receiver operating characteristics (ROC), which gives information about the sensitivity and specificity of each parameter. Within follow-up periods of one, three, six and nine months, PRRI predicted rupture better than both diameter and sex-adjusted diameter. In the same follow-up periods PWRR has predictive quality better than maximum diameter and about the same as sex-adjusted maximum diameter. The exact AUC values can be seen in table \ref{Tab:AUCofPRRIandPWRR}. Other rupture risk indicators besides BRRA can be found in Indrakusuma et al. \cite{Indrakusuma.2016}. \par

\begin{table}[h]
	\centering
	\fontfamily{bch}
	\begin{tabular*}{0.7\textwidth}{c|c|c|c|c}

\textbf{Follow-up period} & \multicolumn{4}{|c|}{\textbf{AUC}}\\
 & \textbf{PRRI} & \textbf{PWRR} & \textbf{Diameter} & \textbf{Diameter (adj.)}\\
$\text{1 month}$ & $0.937$ & $0.905$ & $0.884$ & $0.905$ \\[0.1cm]
$\text{3 month}$ & $0.931$ & $0.897$ & $0.879$ & $0.909$ \\[0.1cm]
$\text{6 month}$ & $0.878$ & $0.859$ & $0.789$ & $0.821$ \\[0.1cm]
$\text{9 month}$ & $0.761$ & $0.727$ & $0.699$ & $0.727$ \\[0.1cm]
$\text{12 month}$ & $0.672$ & $0.606$ & $0.692$ & $0.702$ \\[0.1cm]

	\end{tabular*}
	\caption[AUC values of FEA rupture risk assessments compared to diameter criteria.]{Table gives the AUC of a receiver operating characteristics (ROC) for PRRI and PWRR as FEA-based rupture risk assessments compared to the maximum diameter and the maximum sex-adjusted diameter \cite{Polzer.2020}.}
	\label{Tab:AUCofPRRIandPWRR}
\end{table}

For the evaluation of different model approaches with focus on rupture risk assessment in a clinical environment (see section \ref{subsec:ModelEvalMatrix}), the software system BioPARR \cite{Joldes.2017} and the above mentioned PRRI \cite{Polzer.2015} are utilised as exemplary models based on FEA. 
BioPARR is an open-source software system that allows a rupture risk assessments of AAAs. It was developed to enable a comparative evaluation of AAAs with a standardised approach for researchers and aid clinicians in AAA risk assessment. Besides semi-automatic AAA and ILT segmentation it is fully automated and uses RPI for rupture risk prediction \cite{Joldes.2017}. 
For the rupture risk assessment with PRRI, segmentation and mesh generation is done with the commercially available software A4clinics by VASCOPS GmbH \cite{Polzer.2015}, which is in its pipeline-approach comparable to BioPARR. For FE computation ANSYS (Ansys Inc.) is used. Due the various optimisation algorithms and advanced statistical computation, there is currently no complete software package, that can easily be used in a clinical environment \cite{Polzer.2015}. \par

\textbf{Limitations and open research questions of the FEA approach}

FEA based research has progressed immensely during the last decades and produced numerous results that extended the understanding of aneurysms, especially AAAs. The various biomechanical factors that were researched on highlight the complexity of the disease. But while stand-alone FEA excels in showing the importance of wall stress and strength, it neglects possible influences of the hemodynamic load on AAA mechanisms for growth and rupture. Moreover, FEA is limited to statements based on a model which captures the aneurysm in a static state during the patient's scan. Dynamic processes and developments over time can not be modelled by this approach. With the technological development of increasing computing power, inclusion of CFD and G\&R simulation is possible, towards more complex and realistic models. While this might benefit the research on aneurysm development and rupture mechanisms, it does not necessarily improve patient specific rupture risk assessment or clinical acceptance. Here, long computing times and hardly comprehensible assumptions are seen as disadvantage against simpler decision criteria. This hold true considering the heterogeneity of FEA model in regards to constitutive models, material behaviour and load application. Ongoing uncertainty about the influence of model simplifications on wall stress and strength, diminishes validity of conclusions and makes comparison between computational results difficult. A more standardised approach with definitive guidelines for model creation would be a solution to this uncertainty. Furthermore, new technological or methodological developments towards noninvasive quantification of model parameters like wall thickness, wall compliance and local blood pressure could increase the accuracy of patient specific diagnosis.
Nevertheless, the potential of PWS and related parameter as patient specific rupture risk indicators was demonstrated and they can be of great value for physicians as sole or additional parameters in aneurysm assessment. While different tools for AAA rupture risk prediction in a clinical environment are readily available, research on CAs based on stand-alone FEA models is limited.

\subsubsection{Computational Fluid Dynamics}
\label{subsubsec:CFDstate}
CFD-models are used similar to FEA-models for aneurysm development and rupture probability. While FEA focuses on the wall strength and generally on the structural components of aneurysm pathogenesis, CFD-models give insight into the hemodynamics. As seen in \ref{subsec:AneurysmGrowth}, hemodynamic forces play a central role in the pathogenesis of aneurysms. While this is true for both CAs and AAAs, the role differs for each.\par

\textbf{CFD simulations for CAs} 

Many studies perform CFD simulations to futher the understanding of the different processes taking place in CA development and rupture, as demonstrated in figure \ref{fig:PublicationLocation}. If the focus of the studies lies in a deeper understanding of the processes involved, a generic model of an aneurysm may be sufficient. If the goal is to predict the rupture risk of an CA, a patient-specific model is used. Herein arises a similar challenge as in FEA-models, where a compromise between realistic model and computational effort has to be made. CFD analyses, akin to FEA analyses, involves the processes of imaging, mesh construction followed by computational calculations and post processing. Commonly used imaging techniques to simulate patient specific blood flows are magnetic resonance angiography (MRT), computed tomography angiography (CTA) and 3D rotational angiography (3DRA). Currently, studies suggest that 3DRA is the most prominent technique for patient-specific hemodynamic simulations \cite{Yoon.2016}. \par
For clinical application using CFD as a risk assessment tool, the parameters correlating with rupture risk must be defined. While over the years a large amount of studies have been conducted with the goal of identifying these Parameters, they are still highly debated \cite{Sforza.2012,Xiang.2014b,Jeong.2012c,Murayama.2019}. The parameter most frequently calculated by CFD models is the WSS, which as seen in \ref{sec:ComparisonAAA-CA} plays a key role in aneurysm pathogenesis. Although it is still debated whether high or low WSS is affecting rupture risk dominantly. The definition of WSS varies between different research teams, which makes a direct comparison and validation of different theories difficult. It is worth pointing out though that major studies suggest a correlation of rupture risk with low WSS \cite{Xiang.2011,Xiang.2014b,Detmer.2019,Jing.2015}. Other factors correlating with aneurysm rupture risk include a high area under low WSS \cite{Xiang.2011, Jou.2008,Jing.2015,Zhang.2016} and the oscillating shear index (OSI) \cite{Xiang.2011, Detmer.2019,Jing.2015,Zhang.2016}. The OSI quantifies the degree of WSS oscillation over a cardiac cycle \cite{Tezduyar.2011b}. Higher OSI therefore signals a higher variation of flow direction.
Complex flow patterns, small impingement jet streams and vortex formation seem to also indicate rupture risk \cite{Xiang.2011, Cebral.2005}.\par
It is important to remind that before performing a CFD simulation, its corresponding framework must be defined. Here, the geometry is an important, patient-specific factor \cite{CampoDeano.2015}. The mesh generation needs to be performed with an adequate degree of accuracy, which strongly depends on the present flow condition. For laminar flows, a coarse mesh with fewer cells may be sufficient, whereas turbulent flows require much finer meshes. Most studies anticipate a laminar flow and therefore use relatively coarse meshes, however the true flow condition remains unclear \cite{Saqr.2020}. Besides patient-specific geometries and the mesh size, the fluid behaviour and the boundary conditions have to be defined as well. A still debated controversy is the fluid model assigned to blood. Blood is a non-Newtonian fluid, meaning that viscous stresses cannot be linearly correlated to its strain rate. Nevertheless, around {90}{\%} of CFD simulations model blood as Newtonian fluid \cite{Saqr.2020}. Some studies argue that the differences between Newtonian and non-Newtonian modelling are negligible and therefore Newtonian modelling is preferable due to reduced computing times \cite{Berg.2019b}, while others suggest that a non-Newtonian approach must be applied \cite{Saqr.2020,CampoDeano.2015}. The boundary conditions for the flow simulation have to be defined. At the cross-section of the inlet in the virtual model, a time dependent velocity profile or a waveform is defined. This is either done patient specific, which necessitates measurement of the blood flow or flow rates out of literature. If only the cycle averaged flow field is of interested, a steady flow is simulated. For the cross-section of the outlets, the velocity or pressure is defined via the outlet boundary condition. The most prominent is the zero pressure boundary condition \cite{GalindoRosales.2014}. However it neglects the influence on the pressure and flow field due to downstream effects \cite{Moon.2014}. The increase in computational power enables the use of more realistic outlet boundary conditions, for a detailed discussion see \cite{CampoDeano.2015,Berg.2019b}. 
In simulations the vessel wall is often considered as inelastic, hence the no-slip and no-penetration boundary conditions are applied. For a more realistic approach of vessel wall behaviour, FSI-models can be used to simulated wall deformation over the cardiac cycle. \\
Lastly, the CFD-solver used also impacts the results and must therefore be considered when comparing different studies with each other \cite{Saqr.2020}. For the purpose of validating a CFD-Solver Paliwal et al. \cite{Paliwal.2017} proposed a technique which may be able to act as a future benchmark.
\\ 
For the model comparison performed, CFD based assessment tools for CAs were chosen, which are currently being validated and researched. AView is a Software tool based upon the results of studies executed by Xiang et al. \cite{Xiang.2011,Xiang.2014b,Xiang.2016}. This model correlates risk of rupture with lows WSS and a high OSI, combined with a morphological parameter. While the model itself currently is only validated using a data-set of 204 aneurysms \cite{Xiang.2016}, a pilot software for clinical use was constructed and tested 2017 by 12 different clinicians \cite{Xiang.2017}. Although the participants responded positively to this tool, showing its potential, it is currently not commercially available. As of right now, there are still open questions regarding the proper framework of CFD simulations as a whole and an accepted validation approach.

\textbf{CFD simulations for AAAs}

Looking at the situation in AAA development and risk assessment, CFD-models are limited to further the understanding of AAA formation and development. As elaborated in \ref{sec:ComparisonAAA-CA}, in most AAAs a thrombus is formed, which can provoke a separation between blood flow and vessel wall \cite{Lasheras.2007}. The degeneration of the arterial wall also leads to the destruction of the EC, making the aneurysm wall insensitive to the acting WSS \cite{Vorp.2007}. Therefore, the importance of WSS and other flow specific parameters regarding later stages of AAA growth and rupture is significantly reduced and no tools for risk assessment solely relying on hemodynamics are applied. CFD simulations can be used to asses the hemodynamic load for FEA simulations using FSI models, or calculating the residence time of blood over a cardiac cycle to further the understanding or prediction of thrombus formation.

\subsubsection{Fluid-Structure Interaction}
\label{subsubsec:FSIstate}
Current research acknowledges the importance of the interaction between hemodynamics of the blood flow and stress and deformation of the solid structure, i.e. aneurysm wall and ILT, in regards to its development and rupture prediction \cite{DiMartino.2001, CampoDeano.2015}. Whereas FEA and CFD approaches are unable to model this interaction, FSI simulations couple both computational methods in an attempt to produce more realistic results. 

Modelling techniques for FSI computations are various and complex, because they combine strategies and assumptions of FEA and CFD based approaches. An explanation and comparison of different modelling techniques for CAs is found in Tezduyar et al. \cite{Tezduyar.2011b}, which might also be helpful for AAA modelling. Besides the boundary conditions for FEA and CFD, additional conditions are applied for FSI at the interface between simulated blood flow and solid structure \cite{VALENCIA.2013}:
\begin{enumerate}
 \item The displacement of fluid and wall must be compatible.
 \item The traction must be at equilibrium.
\end{enumerate}
A potential difference between FSI models is their approach on coupling, e.g. the order of calculation between governing fluid and solid equations. While explicit approaches might be sufficient for models with weak fluid-solid interaction, implicit approaches, more precisely iterative implicit or fully coupled, are preferred for increased accuracy. This accuracy comes at the cost of higher requirements for computational memory and simulation time \cite{Salman.2019}.

The use of FSI models and their focus differ by aneurysm type and particular research goal. Most research on CAs based on the FSI approach focuses on the calculation of WSS, while specific statements on circumferential wall stress are omitted or discussed secondary \cite{Torii.2006, Bazilevs.2010, Eken.2017}. The same observation can be made for research on early AAA development \cite{Nestola.2016}. In both, solid structure computation is primarily used to simulate the influence of wall deformation on the fluid and therefore supersede the boundary condition of a rigid wall. This is consistent with earlier observations regarding the focus of literature on CAs on WSS and its general influence on aneurysm initiation and early development. On the other hand, studies on biomechanical risk factors of AAAs mostly use fluid simulation to get a more realistic pressure variation on the aneurysm wall and estimate the principal wall stress more accurately. While research on structural risk factors like the ILT \cite{DiMartino.2001, Rissland.2009} do not account for WSS at all, literature on hemodynamic influences and risk factors \cite{Xenos.2010, Chandra.2016, Drewe.2017, Sharzehee.2018} consider WSS and principal wall stress. The majority of FSI models are used to deepen the knowledge on aneurysm risk factors. Some studies demonstrate its potential in rupture risk prediction for AAAs \cite{Xenos.2015} and CAs \cite{ARANDA.2019} as computational tool to calculate patient-specific wall stress or hemodynamic indicators. Both models couple FSI with an additional approach for rupture risk prediction, namely RPI for AAAs \cite{Xenos.2015} and machine learning for CAs \cite{ARANDA.2019}.

For the evaluation of the model approaches in section  \ref{subsec:ModelEvalMatrix}, the FSI simulation of Aranda et al. \cite{ARANDA.2019} was chosen as representative for the FSI category. The FSI model is utilised to estimated the qualities of different rupture risk indicator on 60 saccular CAs, with an equal number of ruptured and unruptured ones. Several machine learning algorithms are tested on their predictive qualities, based on six significant rupture risk indicators. The research goal is to predict the rupture risk of CAs and aim for a future use in a clinical environment. 

Because FSI simulations require computing power up to multiple orders of magnitude higher than FEA \cite{Leung.2006}, its increase in accuracy compared to CFD and FEA has to be justified. Multiple studies include a comparison with conventional methods examining the difference in WSS or PWS values. There are conflicting results regarding the benefits of FSI on PWS calculation for AAAs. While some research found significant underestimation of PWS values by FEA compared to FSI \cite{Scotti.2008, Chandra.2013}, others observed only minimal difference \cite{Leung.2006, Lin.2017}. Comparison between these studies is difficult, due to varying model assumptions. In the comparison between FSI and CFD, it is indicated that stand-alone CFD overestimates WSS values \cite{Lin.2017}. \par

In summary, FSI models reiterate the patterns seen in the CFD approach, with its focus on the estimation of WSS for CAs, and in the FEA approach, with it centring around PWS calculation for AAAs. Because it combines fluid and solid simulations, FSI models are even more heterogeneous in their assumptions. This complicates a reasonable comparison between distinct studies and results, as seen in the debate about improved accuracy over FEA. While FSI is the more realistic approach compared to CFD and FEA by modelling the interaction between blood flow and solid structure, its benefit in improved accuracy is debated. Nevertheless, there are models that pair FSI with other approaches for rupture risk prediction of CAs and AAAs, that could be utilised in clinics. Concern remains, whether its implementation in a clinical environment is feasible, due to the necessary computing power and long simulation times.

\subsubsection{Fluid-Solid-Growth}
\label{subsubsec:FSGstate}
Humphrey and Taylor \cite{Humphrey.2008} proposed to construct multi-scale models, that combine the interaction between global hemodynamics and local wall stress via FSI and influence of molecular biochemical reactions via growth and remodelling (G\&R). Here, growth describes the change in size and remodelling the change in structure of an aneurysm \cite{Humphrey.2012}. They termed it fluid-solid-growth (FSG) model. While FSI computations capture the state of aneurysms over a cardiac cycle, the G\&R part operates on a timescale of weeks to years to simulate aneurysm development towards either stability or rupture. In an iterative process, wall stress values from the FSI computation are transferred as input to the G\&R simulation, which correlates them to long-term wall deformations via mechanobiological processes. Mechanobiology studies the biological cell response to mechanical stimuli \cite{Humphrey.2012}. The long-term wall deformation influences fluid flow and structural behaviour and is therefore returned to the FSI simulation for the next iteration. As FSG models consider the influence of hemodynamics, structural and biochemical behaviour of the aneurysm wall, the represent the most realistic and complex deterministic modelling approach. In theory, they could be able to model individual aneurysm development from the time of imaging towards an end-stage aneurysm and predict whether it will remain stable or rupture. 

So far, FSG models are utilised quite evenly for AAAs \cite{Sheidaei.2011, Grytsan.2015} and CAs \cite{Ventikos.2009, Watton.2011, Teixeira.2020}. All considered models focus on the mechanobiological response of wall tissue to WSS and pressure, which drives elastin degradation and collagen synthesis. They focus on early aneurysm development with idealised \cite{Ventikos.2009, Watton.2011} or realistic geometries \cite{Sheidaei.2011, Grytsan.2015, Teixeira.2020}. After a bulge is explicitly formed, low WSS drives elastin degradation, which reduces wall distensibility and increases wall pressure. Collagen is synthesised to compensate for increased pressure and stabilise the aneurysm. Synthesis of new wall material drives aneurysm growth \cite{Watton.2011}. A more detailed review on G\&R and FSG models for AAAs can be found in Humphrey et al. \cite{Humphrey.2012}. 
The FSG model for CAs by Teixeira et al. \cite{Teixeira.2020}, representative of the FSG category for the later model evaluation \ref{subsec:ModelEvalMatrix}, adapts of previous FSG models and incorporates the influence of pulsatile hemodynamic indicators like oscillating flow and formation of secondary blebs. Like all other models mentioned above, it does not have a failure criteria that would allow for the aneurysm to develop towards rupture and instead develops it towards homeostasis. \par

While the idea of FSG model and their potential is intriguing, they are far away from any use as tools for rupture risk prediction in a clinical environment. At the moment, they do only account for specific mechanobiological process in regards to elastin and collagen, and omit the influence of SMC behaviour \cite{Grytsan.2015}. The biochemical effect of the ILT, calcifications or general inflammation of the aneurysm wall is not considered. Moreover, these biochemical processes are not fully understood yet, making reliable modelling not possible. As mentioned above, there exists no FSG model that accounts for the possibility of rupture yet. In comparison to FSI, the variety of model assumptions increases for FSG and experimental validation is difficult and so far missing \cite{Humphrey.2012}. This is similarly disadvantageous as the immense need for computing power for a future use in a clinical environment.

\subsubsection{Assessment-tools and dimensionless parameters}
\label{subsubsec:ATDPstate}
With increasing need for accurate models to predict aneurysm rupture, a special focus has been placed on complex and realistic models. Such models may require large computing power, time and investment because the structure and/or the blood flow are simulated. For critical situations where time is of essence, such powerful and accurate models may take too much time or be generally inaccessible to hospitals. 

Hence the need for faster models such as assessment-tools and dimensionless parameters (AT\&DP) that predict growth and aneurysm rupture arises. Assessment-tools list factors whose data is required from the patient. Each factor has been included in the list after much research such as the frequency and range of values of said factor when investigating aneurysm rupture. Furthermore, every factor has a specific weight which corresponds to its importance in predicting aneurysm rupture. Dimensionless parameters encompass equations which classify aneurysms from aneurysm rupture. These equations are derived from more complex models such as CFD and FEA. 

In the scope of this paper, three models have been considered. The chosen models are either presently used in clinics or show a high potential as a possible tool in the future. These are Aneurysm number (An), UIATS (Unruptured Intracranial Aneurysm Treatment Score) and PHASES (Population, Hypertension, Age, Size, Earlier subarachnoid hemorrhage, and Site). ELAPSS (CA) \cite{vanSanchezKammen.2019} \& the Writhe Number (CA) \cite{Lauric.2011} were identified as two more promising AT\&DP approaches. Nevertheless, since they are seldom used in clinics and hospitals, they have not been further discussed in this paper.
Aneurysm number is a dimensionless parameter as described by Asgharazadeh et al. \cite{Asgharzadeh.2019} to detect CA, more specifically sidewall and bifurcation aneurysms. This dimensionless parameter is calculated by multiplying the ratio of aneurysm width (\textit{L}) and artery diameter (\textit{D}) with the respective pulsatile index (\textit{PI}) and $\alpha $ (1 for sidewall and 2 for bifurcation).

\begin{equation}
    An = \alpha \frac{L}{D} PI
    \label{eq:AneurysmNumber}
\end{equation}
\begin{equation}
    PI = \frac{\Delta U}{U}
    \label{eq:PulsatileIndex}
\end{equation}
\begin{equation}
    \Delta U = {U_{max}} - {U_{min}} 
    \label{eq:DeltaU}
\end{equation}

Pulsatile index, also known as Gosling’s pulsatile index, is the ratio of the difference between maximum and minimum velocities during the cardiac cycle in the parent artery to the average velocity (\textit{U}) in the parent artery. Aneurysm geometric measurements can be calculated from medical images of the aneurysm with clear concise methods, while \textit{PI} can either be calculated or extracted from literature \cite{Ali.2013}. When An is less than 1, it depicts cavity flow mode of the blood which represents lower rupture risk. Vortex mode is represented when An is greater than 1, representing higher rupture risk which correlated with higher wall-shear stress (WSS) and oscillatory shear stress (OSS).

Furthermore this dimensionless model is also implementable with the help of machine learning algorithms. As mentioned in \cite{Asgharzadeh.2020}, their model had achieved an AUC of 0.90 which shows the high prediction quality in detecting between higher and lower rupture risk. These results depict the potential of the Aneurysm Number in predicting rupture risk, especially when incorporated with machine learning models. Furthermore application of \textit{An} can be further improved by combining it with statistical or ML methods which makes use of regression analysis to “fit” \textit{An} number with the rupture rate \cite{Asgharzadeh.2020}. 

Another assessment tool with promising performances is UIATS, which was invented by 69 specialists to develop a concise yet effective assessment tool for dealing with CAs \cite{Etminan.2015}. It accounts for 29 key factors (age, risk factor incidence, clinical symptoms, life expectancy due to chronic/malignant disease, comorbid disease, maximum diameter, morphology, age-related risk, aneurysm size-related risk, aneurysm complexity-related risk), when advising against or for treatment of unruptured CAs. The scores for UIA-repair and conservative approach are calculated independently from each other. Conservative approach advises against operative treatment and advises change in habits and/or use of medicines. This is done with the help of factor-specific defined point systems. If the score difference between the two approaches is more than 3, the approach with the highest points is then chosen. However if the score difference is 2 or less than 2, then the UIATS recommends either approach while considering other factors.
A couple of validation studies were conducted in the US, where unruptured intracranial aneurysms (UIAs) were observed in 221 patients \cite{Ravindra.2018}, 147 UIAs in Germany \cite{HernandezDuran.2018} while 71 UIAs in Spain (71 UIAs) \cite{Mateo.2018}. The study in US concluded that the UIATS generally recommended overtreatment of UIAs, and proposes to use the tool rather as a screening tool \cite{Ravindra.2018}. Contrary to that, independent studies in Germany and Spain indicated a positive correlation between actual treatment and the treatment proposed by the UIATS \cite{HernandezDuran.2018} \cite{Mateo.2018}. However, it must be used with caution so as to avoid unnecessary treatment when the risks are low to non-existent. 

PHASES score is an assessment tool taking six factors (population, hypertension, age, size of aneurysm, earlier subarachinoid hemorrhage and site of aneurysm) into consideration to analyse CAs. The total points calculated correspond to a 5-Year absolute risk of rupture in percentage \cite{Greving.2014}. Patient-information and the diameter of the aneurysm are required to predict aneurysm rupture. Higher PHASES scores correlate with higher rupture risk and increased aneurysm growth. PHASES score is already used commonly in clinics and hospitals to determine the absolute risk of rupture and has, since its conception, been validated and proven by multiple studies \cite{Mateo.2018}. However, PHASES score has been deemed to underrepresent, patients with familial aneurysms or patients who are young smokers \cite{etminan2015unruptured}. 

Overall more AT\&DP exist for CAs than for AAAs. However, the diameter criterion is still used when determining the rupture risk of AAAs. There are other possibilities to use the wall thickness or derived values for assessing rupture risk. AT\&DP offer various benefits especially when time is of the essence. These tools do not require highly trained medical personal or experts to be applied, since the majority of the required data is either data provided by the patient himself such as age, habits (e.g. smoking, drug-abuse) or ethnicity, or the data can be extracted directly from common medical imaging techniques. 

The easy availability of data reduces the time necessary to estimate or calculate the aneurysm rupture risk and consequently reduces the cost in utilising these models. In general, assessment-tools are user-friendly and ready-to-use in clinics and hospitals. Although AT\&DP are more accurate than diameter criterion, they are less accurate when compared to FEA and CFD which are patient-specific models. Furthermore, AT\&DP are specific to certain types of aneurysm and are only efficient if its limitations are taken into consideration.

\subsubsection{Machine Learning}
\label{subsubsec:MLstate}
Machine learning is especially for complex problems with many influence parameters and complex interactions useful, problems that can not be easily explained with white box models. Since the mechanisms occurring in aneurysms and their interactions regarding pathogenesis are not fully understood, ML seems to be a promising approach, as it can be applied as grey or black box model \cite{pintelas.2020}. 

Input data for Machine Learning models can be categorised into clinical images, clinical patient data (sex, age, weight, ...), morphological parameters and hemodynamic or structural parameters. If hemodynamic or structural parameters are used as input, ML models are usually applied in combination with CFD or FEA models. Machine Learning models are applied in different contexts for aneurysms. There exist ML algorithms for image analysis and segmentation, rupture risk evaluation, growth prediction, estimation of post operational mortality, WSS estimation and for parameter extraction of CFD/FEA simulations. 

Technologically advanced models extract morphological features automatically from clinical images. Image analysis and segmentation then is combined with rupture risk evaluation in one model. Clinical images analysed by an algorithm are inherently more neutral and not affected by inter-observer variability compared to doctors \cite{Parikh.2018, Rengarajan.2020b}. In general, to identify ML approaches with potential of being applied for clinical aneurysm rupture risk evaluation, the focus was set on models including over 100 aneurysms that therefore have a relatively high significance,
models that are highly automated and models that showed good performance metrics on test and evaluation data. In addition, the models should be newer than 4 years to incorporate current advances in computer science. Models combining ML approaches with features extracted from FEA or CFD simulations have high potential to further improve prediction performances by generating synergies from both approaches. Doing so, geometries and boundary conditions could be extracted automatically from clinical images using an ANN, this information again is used to set up and run a CFD or FEA simulation. Parameters calculated by these simulations such as for example WSS, WSSG or PWS can afterwards be used as additional features for another ML algorithm to finally evaluate the underlying rupture risk. However, such combined models are considered as either FEA or CFD approaches in this review.

Based on the aspects outlined, one ML state of the art model for each of the aneurysm types (CA and AAA) was identified to be further evaluated in section \ref{sec:Evaluation}:

Zhu et al. \cite{Zhu.2020} developed ML models with several different algorithms using clinical and morphological features to perform a stability assessment for cerebral aneurysms. The model using ANN showed best performance with an accuracy of 0.824 and an AUC of 0.867. The study regarded 18 morphological and 13 patient-specific features of a total of 2067 aneurysms. Geometric features are calculated using GEOMAGIC 12.0 and the risk assessment is afterwards performed using an ANN engineered with MATLAB. This model only works for saccular cerebral aneurysms but the concept of automatic geometry extraction and afterwards risk assessment using geometrical and patient-specific features could be transferred for abdominal aneurysms, too. 

Parikh et al. \cite{Parikh.2018} developed a decision tree model to be used to support aneurysm rupture risk classification based on geometry. Nevertheless, this algorithm does not directly evaluate the rupture risk but categorizes aneurysms as either electively or emergently repaired. Since aneurysms that undergo emergently repair are confronted with a high rupture risk, the relevance of this study is given. AAA centerline length, L2-Norm of Gaussian curvature and Wall surface area were identified as most important features. In total 150 AAA CT-scanned by two different hospitals out of which 75 were emergently and 75 were electively repaired were regarded. Using an in-house segmentation code the aneurysm is segmented into lumen, inner wall and outer wall. Afterwards, contours are identified that represent the aorta outer wall boundary and the best one is selected by the user. For the performed classification an accuracy of 0.955 and an AROC of 0.96 was realised by the Decision Tree. The algorithm considers so far only geometry for classification. To adapt it for rupture risk prediction accordingly, using clinical parameters might further improve the results. 

Although different researchers use different ML algorithms for their specific sets of data and features, they face similar challenges, no matter whether they model CAs or AAAs. 
Because Machine Learning is closely related to statistics, the model predictions are only statistically correct. It is consequently relatively unlikely that a ML algorithm evaluates a small AAA with a high rupture risk. To get better prediction results applicable for clinical use, there is a need for larger databases existing out of both, clinical images and patient-specific data \cite{Shi.2020, Park.2019, Parikh.2018} to better capture underlying stochastic patterns. To establish such a database questions concerning related privacy issues and image rights need to be solved \cite{Raffort.2020}. For the creation of large databases involved economical aspects must be discussed \cite{Raffort.2020, Shi.2020} and common standards for clinical images should be set to assure risk evaluation algorithms can be applied in every hospital without restricting them to a specific scanner manufacturers or similar \cite{Raffort.2020, Zhu.2020}. Training times for ML algorithms are outlined in literature as another challenge \cite{Raffort.2020, Zhu.2020} but comparing it to computation times of patient-specific CFD or FEA simulations, it is only a minor challenge. Nevertheless, current ML models for CAs are limited on saccular shapes \cite{Zhu.2020, Chen.2020} and models for AAAs are limited on a fusiform shape \cite{Zhu.2020, Chen.2020}.

Successful application of ML tools in clinical aneurysm risk evaluation can only be established with interdisciplinary collaboration between computer scientists, radiologists and clinicians. Since AAAs and CAs are confronted to similar problems, the establishment of a mutual ML approach that is afterwards adapted for each aneurysm type's peculiarities should be discussed. Making these adaptions, research outcomes from other modelling approaches and earlier research is crucial. Therefore, for example AAAs must be modelled taking ILTs into account (\cite{Meng.2014, Frosen.2012}) and CA modelling must consider for blebs \cite{Cebral.2010} and the specific cerebral location \cite{Zhu.2020}. 

All in all, ML is a highly promising modelling approach for aneurysms that successfully demonstrated its potential after being in scientific focus for only a relatively short time \cite{Rengarajan.2020b}.

\section{Model Evaluation}
\label{sec:Evaluation}
This section builds upon the results of the systematic model review and evaluates the six different model approaches regarding their ability in rupture risk prediction and applicability in a clinical environment. In section \ref{subsec:ModelEvalMatrix}, at least on exemplary model of each category is chosen and assessed, based on six evaluation criteria. Following in section \ref{subsec:GenEvalOfModelApproach}, the results from the exemplary models and from the systematic model review are combined to give a conclusive evaluation of the modelling categories. Lastly, in section \ref{subsec:Future} and \ref{subsec:Limitations} possible future directions and the limitations of this study are discussed. CFD, FEA, FSI and FSG models are broadly categorised as deterministic models, whereas ML and AT\&DP models are categorised as stochastic models.

\subsection{Model Evaluation Matrix}
\label{subsec:ModelEvalMatrix}
In order to compare the various models for aneurysm rupture risk prediction, a matrix for model evaluation, shown in table \ref{Tab:ModelOverview} was designed. Here, 11 models are evaluated on six different criteria, listed below. Each model category considered in section \ref{subsec:StateOfTheArt}, is represented by at least one model. Their rating for each criterion can ranges from -- (-2 points), via - (-1 point), 0 (0 point), and + (+1 point) to ++ (+2 points). These ratings are then multiplied with the “weight of importance” of each evaluation criterion. The weights of importance were identified using the Analytic Hierarchy Process (AHP) \cite{Coyle.2004}. 

The AHP is a method to structure subjective decision making (here calculation of weight of importance) between various options (here evaluation criteria). It utilises a pairwise comparison between two options on ones importance over the other. The importance is rated on a scale between 1 and 9 (see Saaty Rating scale in appendix \ref{S3_Fig}). Each rating was assigned in a discussion between four group members. Based on these ratings, an eigenvector is calculated which gives the relative importance of each criteria, e.g. the weight of importance. The AHP contains a method to check for consistency of the ratings, the consistency ratio (CR).

The following list describes the criteria with their corresponding “weight of importance”:
\begin{itemize}
\item \textbf{Model Complexity (4.98\%)} \\ Considers time taken to set up the model as well as the internal complexity.
\item \textbf{Readiness Level (26.69\%)} \\ Evaluates the maturity of the model in regards to its use in a clinical environment.
\item \textbf{Prediction Quality (44.27\%)} \\ Compares the prediction result to the frequently used diameter criterion, where the diameter of the aneurysm is the deciding factor in predicting aneurysm rupture and whether an operation is deemed necessary.
\item \textbf{Cost (3.27\%)} \\ Considers for capital costs and running costs.
\item \textbf{User-friendliness (9.52\%)} \\ Evaluates whether pre-knowledge is necessary to apply the model and the ease with which the software can be learned, used and understood. 
\item \textbf{Model limitations (11.28\%)} \\ Considers the application of the model for different aneurysm types. Additionally, it includes limitations that are predominant in the respective modelling category.

\end{itemize}

Conclusively, the total points are calculated by multiplying the ratings for each aspect reaching from -- to ++ with the “weights of importance” and adding them up. The achievable rating scale is standardised so that each model is finally rated with points ranging from 0 to 100. Table \ref{Tab:ModelOverview} shows the suggested model evaluation matrix with an evaluation performed for 11 different models. 

\begin{table}[h]
\begin{adjustwidth}{-2.4in}{0in} 
	\centering\fontfamily{bch}
	\begin{tabular*}{1.3\textwidth}{@{}l@{\extracolsep\fill}l@{\extracolsep\fill}l@{\extracolsep\fill}l@{\extracolsep\fill}l@{\extracolsep\fill}l@{\extracolsep\fill}l@{\extracolsep\fill}l@{\extracolsep\fill}@{\extracolsep\fill}l@{\extracolsep\fill}l@{\extracolsep\fill}l@{\extracolsep\fill}l@{\extracolsep\fill}}
\textbf{}&																			\textbf{Model 1} &
\textbf{Model 2} &
\textbf{Model 3} &
\textbf{Model 4} &
\textbf{Model 5} &
\textbf{Model 6} &
\textbf{Model 7} & 
\textbf{Model 8} &
\textbf{Model 9} &
\textbf{Model 10} &
\textbf{Model 11} \\

$\text{Model Name}$&	
$\text{C5.0 }$ & 
$\text{ANN}$ &
$\text{BioPARR}$ &
$\text{Probab.}$ &
$\text{Integrative}$ & 
$\text{AView}$ &
$\text{}$ & 
$\text{}$ & 
$\text{Aneurysm}$ &
$\text{UIATS}$ &
$\text{PHASES}$ \\ &
$\text{Decision}$ &
$\text{}$ &
$\text{}$ &
$\text{Rupt. Risk}$ &
$\text{Mech.-bio.}$ & 
$\text{}$ &
$\text{}$ &
$\text{}$ &
$\text{Number}$ &
$\text{}$ &
$\text{}$ \\  &
$\text{Tree}$ &
$\text{}$ &
$\text{}$ &
$\text{Index}$ &
$\text{Framework}$ &
$\text{}$ &
$\text{}$ &
$\text{}$ &
$\text{}$ &
$\text{}$ &
$\text{}$ \\  [0.1cm]

$\text{Author}$&	
$\text{Parikh}$ & 
$\text{Zhu}$ &
$\text{Joldes}$ &
$\text{Polzer}$ &
$\text{Teixeira}$ &
$\text{Xiang}$ &
$\text{Detmer}$ &
$\text{Aranda}$ &
$\text{Asgharzadeh}$ &
$\text{Etminan}$ &
$\text{Greving}$ \\ & 
$\text{et al.}$ &
$\text{et al.}$ &
$\text{et al.}$ &
$\text{et al.}$ &
$\text{et al.}$ &
$\text{et al.}$ &
$\text{et al.}$ &
$\text{et al.}$ &
$\text{et al.}$ &
$\text{et al.}$ &
$\text{et al.}$ \\  [0.1cm]

$\text{Year Published}$&	
$\text{2018}$ & 
$\text{2020}$ & 
$\text{2017}$ & 
$\text{2015}$ & 
$\text{2020}$ & 
$\text{2017}$ & 
$\text{2018}$ & 
$\text{2019}$ & 
$\text{2020}$ & 
$\text{2015}$ & 
$\text{2014}$ \\ [0.1cm]

$\text{Reference}$&	
$\text{\cite{Parikh.2018}}$ & 
$\text{\cite{Zhu.2020}}$ & 
$\text{\cite{Joldes.2017}}$ & 
$\text{\cite{Polzer.2015}}$ & 
$\text{\cite{Teixeira.2020}}$ & 
$\text{\cite{Xiang.2017}}$ & 
$\text{\cite{Detmer.2019}}$ & 
$\text{\cite{ARANDA.2019}}$ & 
$\text{\cite{Asgharzadeh.2019}}$ & 
$\text{\cite{Etminan.2015}}$ & 
$\text{\cite{Greving.2014}}$ \\ \\[0.1cm]

$\text{Approach}$&	
$\text{ML}$ & 
$\text{ML}$ & 
$\text{FEA}$ & 
$\text{FEA}$ & 
$\text{FSG}$ & 
$\text{CFD}$ & 
$\text{CFD}$ & 
$\text{FSI}$ & 
$\text{AT\&DP}$ & 
$\text{AT\&DP}$ & 
$\text{AT\&DP}$ \\[0.1cm]

$\text{Location}$&	
$\text{AAA}$ & 
$\text{CA}$ & 
$\text{AAA}$ & 
$\text{AAA}$ & 
$\text{CA}$ & 
$\text{CA}$ & 
$\text{CA}$ & 
$\text{CA}$ & 
$\text{CA}$ & 
$\text{CA}$ & 
$\text{CA}$ \\[0.1cm]

$\text{Imaging}$&	
$\text{CT}$ & 
$\text{DSA}$ & 
$\text{CT, MRI}$ & 
$\text{CT}$ & 
$\text{CT}$ & 
$\text{CT}$ & 
$\text{CT}$ & 
$\text{CT}$ & 
$\text{CT}$ & 
$\text{CT}$ & 
$\text{CT, MRI}$ \\ \\[0.1cm]

$\textbf{Evaluation:}$ \\[0.1cm]

$\text{Model Complexity}$&	
$\text{+}$ & 
$\text{0}$ & 
$\text{+}$ & 
$\text{- -}$ & 
$\text{-}$ & 
$\text{-}$ & 
$\text{-}$ & 
$\text{-}$ & 
$\text{++}$ & 
$\text{+}$ & 
$\text{++}$ \\ [0.1cm]

$\text{Readiness Level}$&	
$\text{0}$ & 
$\text{+}$ & 
$\text{+}$ & 
$\text{++}$ & 
$\text{- -}$ & 
$\text{++}$ & 
$\text{0}$ & 
$\text{0}$ & 
$\text{0}$ & 
$\text{+}$ & 
$\text{++}$ \\[0.1cm]

$\text{Prediction Quality}$&	
$\text{+}$ & 
$\text{++}$ & 
$\text{+}$ & 
$\text{++}$ & 
$\text{0}$ & 
$\text{+}$ & 
$\text{+}$ & 
$\text{0}$ & 
$\text{+}$ & 
$\text{0}$ & 
$\text{+}$ \\[0.1cm]

$\text{Cost}$&	
$\text{+}$ & 
$\text{+}$ & 
$\text{-}$ & 
$\text{- -}$ & 
$\text{-}$ & 
$\text{0}$ & 
$\text{-}$ & 
$\text{-}$ & 
$\text{++}$ & 
$\text{++}$ & 
$\text{++}$ \\[0.1cm]

$\text{User-friendliness}$&	
$\text{+}$ & 
$\text{++}$ & 
$\text{+}$ & 
$\text{-}$ & 
$\text{-}$ & 
$\text{-}$ & 
$\text{- -}$ & 
$\text{- -}$ & 
$\text{+}$ & 
$\text{++}$ & 
$\text{++}$ \\[0.1cm]

$\text{Model limitations}$&	
$\text{+}$ & 
$\text{+}$ & 
$\text{-}$ & 
$\text{+}$ & 
$\text{++}$ & 
$\text{0}$ & 
$\text{0}$ & 
$\text{+}$ & 
$\text{-}$ & 
$\text{-}$ & 
$\text{+}$ \\[0.1cm]

$\textbf{Points}$&	
$\text{68}$ & 
$\text{87}$ & 
$\text{68}$ & 
$\text{82}$ & 
$\text{38}$ & 
$\text{71}$ & 
$\text{54}$ & 
$\text{46}$ & 
$\text{65}$ & 
$\text{61}$ & 
$\text{86}$ 

	\end{tabular*}
	\caption{Overview Model Evaluation}
	\label{Tab:ModelOverview}
	\end{adjustwidth}
	\end{table}

\subsection{General Evaluation of Modelling Approaches}
\label{subsec:GenEvalOfModelApproach}
The model evaluation matrix serves as an exemplary platform for a comparison of the different modeling approaches discussed in \ref{subsec:StateOfTheArt}. The 3 criteria with the largest weights are prediction quality ({44.27}{\%}), readiness level ({26.69}{\%}) and model limitations ({11.28}{\%}). They amount to a total of {82.24}{\%} and therefore play a dominant role in the model evaluation. Nevertheless, all criteria should be considered and their importance may vary depending on the specific purpose of the model. This section discusses the strengths and weaknesses of the various modelling approaches, based on the mentioned criteria and represented by the exemplary models in the evaluation matrix. \par

Model complexity focuses on the necessary time for rupture risk assessment with the each model. Deterministic approaches are at an inherent disadvantage, since they rely on solving numerical equations, which involves long computation times. This especially reigns true for more complex approaches like FSI and FSG. As discussed in \ref{subsec:StateOfTheArt}, there is a tendency towards more complex models, which consider the interaction between multiple influencing processes. This may lead to longer assessment times, especially if the complexity increases faster than new advances for more powerful computing hardware. AT\&DP models provide a fast rupture risk evaluation, because they rely solely on patient and geometrical data. This explains the high performance of Model 9 to Model 11 under this criteria. This model approach should be considered when time is of essence. ML approaches score in between deterministic approaches and AT\&DP, as they depend on the available computing power, but usually have a low assessment time, once the algorithm is fully trained. \par

A detailed overview of the current readiness level for each modelling approach is given in section \ref{subsec:StateOfTheArt}. Complex models like FSI and FSG perform the worst in this regard, which is mainly due to the fact that they are comparatively new, not as frequently used as FEA and CFD models and a common framework is still debated. While ML models are fairly new as well, they have the advantage of being a grey or black box model, hence a deep understanding of the mechanisms involved is not necessary for rupture risk evaluation and no model assumptions have to be made. Exemplary models for FEA (3 and 4) and CFD (6, but not 7) score high in this category, because they are build as tools that are ready to be implemented in clinics. While CFD and FEA share a long history in aneurysm research, fluid-based models in general are still considered to be less developed for utilisation in a clinical environment. Reasons for this are the fewer numbers of agreed factors correlating with rupture risk compared to FEA. Furthermore there are newer studies challenging common model assumptions, for example the Newtonian or non-Newtonian behaviour of blood. Therefore the high performance of model 6 under this criteria is atypical for the CFD modelling approach in general. Considering these points, it is not unlikely that ML models will be widely accepted by clinicians before CFD models are. AT\&DP tools as a whole perform best in readiness level since assessment tools are already in use for aneurysm rupture risk prediction and other diseases. For example, the PHASES Score (Model 10) is already being used as a risk assessment tool for evaluating CAs in the clinical environment. \par 

Prediction quality is of major importance and has the largest weight on the total score. Both, to recommend an intervention for stable aneurysms and further monitor aneurysms on the brink of rupture involves high risk of unnecessary and possibly fatal consequences. Most models for rupture risk prediction are validated with a ROC and compared to the diameter criterion. Whether the model is of deterministic or stochastic nature is of little relevance, as both approaches have the potential to outperform the diameter criterion. Therefore the criteria of prediction quality is preferably evaluated model by model. In general, a certain rupture risk prediction model should only be considered for clinical use if it performs better than the diameter criterion. A score of 0 is given for models with similar prediction qualities (model 10) and modelling approaches that have not reached the readiness level for rupture risk assessment (model 5 and 8). The ML-based model 2 and the FEA-based model 4 lead the comparison for prediction quality, because they do not only outperform the diameter criterion but were also validated under more reliable conditions than other represented models. \par

Regarding cost, deterministic modelling approaches perform worst. They require expensive computing hardware and often additional software purchases. Besides these investment costs, additional staff may be hired and trained to perform these complex simulations. Additionally, software licensing might add to the running costs. For ML-based models, required computing power is lower than for deterministic approaches, but they require a database to train the algorithms. AT\&DP tools perform best in this category, because they come with little to no additional costs associated with them. They only rely on patient and geometrical data that is already acquired when the patient is examined. \par

For a model to be used for rupture risk assessment, its information has to be conveyed to the responsible clinician in an comprehensible manner. As deterministic modelling approaches are mostly developed by engineers, they may rely on knowledge which is not common in the medical community. It is unlikely that the user, in this case the clinician, is aware about the implications that a specific assumption of the model framework might have on the rupture risk assessment. The adaption to the user should be carefully considered, when designing a simulation-based modelling approach. While deterministic approaches score low in this category, positive examples for adaption to the user can also be found (model 3). Stochastic models perform better, especially AT\&DP models, such as the PHASES score. They are already in use for aneurysm risk assessment, and these techniques are well-established within the medical field. Henceforth new types of AT\&DP models may be easily implemented into the clinical work field. ML-based models are new and evolving quickly due to the increase in computing power. Consequently they have not been established in a clinical environment. However, as grey or black-box models they may not require a deep understanding of the used algorithm and its assumptions compared to deterministic models.\par

Lastly, model limitations depend largely on the model being discussed. In general, they share similarities independent of the used modelling approach. Models for AAA rupture risk assessment focus primarily on fusiform geometries, whereas those for CAs focus on an saccular shape. As examined in section \ref{subsubsec:MLstate}, ML models are limited by the number and variety of aneurysms used to train the model. They are currently only able to adequately predict the risk of larger aneurysms. Deterministic approaches have advantage regarding model limitations compared to stochastic models. Because they consider the underlying mechanisms influencing aneurysm rupture, they can account for unique occurrences in an individual aneurysm. More complex FSG models seem to be most promising in this regard as they account for most processes involved in aneurysm development. With ongoing research, further knowledge and consideration of biomechanical factors for rupture risk limitations for deterministic and stochastic models will decline. 

With above 80 points, the ML-based ANN (model 2), the FEA-based PRRI (model 4) and PHASES (model 11) score the highest in the model evaluation matrix. While the stochastic models ANN and PHASES consider the rupture risk of CAs, PRRI as a deterministic model is used to assess the rupture risk of AAAs. This emphasises the difference in utilisation of modelling approaches between CAs and AAAs. The deterministic approach mostly used for CAs, namely CFD, lacks answers for multiple questions regarding biomechanical influences, model assumptions and validation, as discussed in section \ref{subsec:StateOfTheArt}. As CFD-based models are unable to reliably predict CA rupture risk, stochastic approaches like ML and AT\&DP are therefore tried instead. While PHASES is already in use in a clinical environment, upcoming ML models shows promise to be implemented in clinical rupture risk assessment in the future. On the other hand for AAAs, there seems to be more consensus on basic research compared to CAs, as well as more advanced deterministic FEA models for rupture risk assessment. Hence, various FEA-based tools, like PRRI, are available to be tested and implemented in a clinical environment. Nevertheless a significant breakthrough as a widespread replacement or addition to the often criticised diameter criterion has not happened yet. Reasons might be the high model complexity and difficulties in providing sufficiently user-friendliness. 

Currently, combined deterministic models for both CAs and AAAs used for rupture risk assessment seem not feasible. While sole CFD or FEA based models do not account for all characteristic mechanisms of each aneurysm type, FSI and FSG approaches might be able to consider all relevant interactions, but have not yet reached the readiness level required for a clinical application. 

\subsection{Future Perspectives for Rupture Risk Assessment}
\label{subsec:Future}

At the moment the most promising approach seems to be a combination between deterministic and stochastic models for rupture risk prediction. Such grey-box models are able to benefit from strengths of both categories, namely the fewer model limitations of deterministic approaches with the better user-friendliness and lower complexity of stochastic models. Parameters calculated with FEA or CFD models can be included into ML and AT\&DP approaches, increasing their respective prediction quality. Especially ML approaches seem to be promising, due to the ongoing increase in available computing power and a scalable prediction quality that depends only on the available data for algorithm training.
For these grey-box models, it is important to balance the complexity of the simulation with consideration for the needs of the user and thorough validation, if they are supposed to find wide acceptance in a clinical environment. 

Deterministic approaches, especially complex FSI and FSG, are highly important for future research on mechanisms and biomechanical factors involved in aneurysm formation, development and rupture. With further research insights and consensus on biomechanical factors involved in aneurysm rupture, relevant parameters improve also the prediction quality of stochastic and combined grey-box approaches. 

Overall, aneurysm modelling approaches can not replace doctors in clinical aneurysm management but must be considered as tools to support and improve their decisions. Algorithms and simulations can not consider personal factors of the patient's environment that can be highly relevant for aneurysm treatment recommendations as well.

During this research, it was noticed the absence of an approach to model an aneurysm's development towards rupture similarly to the concept of fatigue in material science, e.g. the weakening of the material under cyclic load. Solely Avolio et al. \cite{Avolio.1998} examined the behaviour of elastin under cyclic load and in relation to the total number of cardiac cycles of different mammals. An adaptation of the fatigue concept to aneurysms could model rupture as a result of progressing local weaknesses, instead of a sudden event, when wall stress exceeds wall strength. Hereby, the energy input into the arterial wall during each cardiac cycle could be considered as driving force for aneurysm progression. Research into this idea seems to be a promising alternative approach on aneurysm rupture.

\subsection{Limitations of this Work}
\label{subsec:Limitations}
There are certain limitations that are prevalent in this review and should be considered together with its results:\par

The creation and evaluation of exemplary models in section \ref{subsec:ModelEvalMatrix} is biased in regards to which models have been chosen from each category, how they are rated and how each criteria is weighted. Even though the structured AHP method was utilised and consensus between the authors was reached, the calculated "weights of importance" are biased due to their personal experience. People with a different background may rate the criterias with a different importance, which would influence the final score of each model. For these reasons, the final score should only be considered as rough indicator and small differences between models have little significance. 

Furthermore, the decision to consider only reviews out of all literature found during the systematic search, harbours the possibility that specific papers might have been missed, if they were not picked up by any review. To account for this, additional searches have been made for each modelling approach to screen for relevant literature with a focus on recent publications.

\section{Conclusion}
\label{chap:Conclusion}
At last, the focus is brought back to the initial research questions of this paper. In section \ref{sec:ComparisonAAA-CA}, cerebral and abdominal aortic aneurysms were compared. While there are inherent differences in the pathogenesis and development of CAs and AAAs, there are also similarities. As shown, both are heavily influenced by WSS and inflammatory processes leading to structural degeneration of the vessel wall. Therefore ongoing collaboration between interdisciplinary teams should be promoted, to benefit from each others knowledge and advance research on this complex disease. \par
    
Additionally, an overview over the current state of different modelling approaches for aneurysm rupture risk assessment was given with the focus on clinical utilisation for CAs and AAAs. As of currently, deterministic approaches, mainly CFD, FSI and FSG lack behind the stochastic approaches of ML and AT\&DP. With a trend towards increasingly more complex models, it seems unlikely that deterministic tools, besides FEA approaches for AAAs, will be used by clinicians in the near future. However, these complex deterministic tools may lead to a better understanding of processes and risk factors for aneurysm development and rupture. Therefore, they can contribute to rupture risk assessment, especially for AT\&DP and ML approaches. In addition, a combination between comprehensible, thoroughly validated deterministic and powerful stochastic models shows potential to positively influence the prediction quality and model limitations compared to models solely relying on a stochastic approach. Models incorporating both CAs and AAAs seem currently not feasible for rupture risk assessment in a clinical environment.

\section*{Supporting information}

\paragraph*{S1 Table}
\label{S1_Fig}
{\bf Search Terms for the general literature research.}

\begin{table}[h]
\begin{adjustwidth}{-2.25in}{0in} 
	\centering\fontfamily{bch}
	\begin{tabular*}{0.8\textwidth}{@{}l@{\extracolsep\fill}p{11cm}@{\extracolsep\fill}}

\textbf{Search Aspect}&											
\textbf{Search term} \\
All Data &										
TI  =  (aneurysm*)  \\
		&  AND  TI=(simulation* OR  model*  OR  index* OR  parameter* \\
		&	{OR  machine  learning*  OR  artificial  intelligence*  OR  decision  tree*} \\
		&	{OR  random  forest*  OR  data  mining* OR  neural  network* } \\
		&	{OR  deep  learning*  OR  Computational* OR finite  element*} \\
		&	{OR  Rupture  Risk  OR  Biomechanic*  OR  structur*  OR  fluid*} \\
		&	{OR  FSI* OR  FSG* OR solid OR G\&R OR Remodeling)} \\
		&	{NOT  TI=(rabbit* OR  animal*  OR  stent*  OR  mouse* OR  pig*}\\
		&	{OR  porcine*  OR  rat*  OR  in  vivo*  OR  mice*  OR  economic} \\
		&	OR  murine* OR  Coil*  OR  Printing*  Or  Clip*  OR  dog  OR  canine)\\ [0.1cm]
\end{tabular*}
	\caption{Web of Science search terms for All Data of Meta-Analysis}
	\label{Tab:Datenbankabfrage1}
	\end{adjustwidth} 
	\end{table}

\paragraph*{S2 Table}
\label{S3_Fig}
{\bf AHP matrix.}

\begin{table}[ht]
\begin{adjustwidth}{-2.25in}{0in} 
	\centering\fontfamily{bch}
	\begin{tabular*}{0.8\textwidth}{@{}l@{\extracolsep\fill}p{4.7cm}l@{\extracolsep\fill}p{7cm}@{\extracolsep\fill}}

\textbf{Intensity of importance }&	\textbf{Definition} & \textbf{Explanation}\\
$1$&												Equal importance &	Two factors contribute \\
            & & equally to the objective \\[0.1cm]
${3}$& Somewhat more important &Experience and judgement    \\
            & & slightly favour one over \\
            & & the other \\[0.1cm]
${5}$& Much more important  &	Experience and judgement   \\
            & & strongly favour one over \\
            & & the other \\[0.1cm]
${7}$& Very much more important &Experience and judgement \\
            & & very strongly favour one \\
            & & over the other. Its importance \\
            & & is demonstrated in practice. \\[0.1cm]
${9}$& Absolutely more important &The evidence favouring one  \\
            & & over the other is of the \\
            & & highest possible validity \\[0.1cm]
${2,4,6,8}$& Intermediate values&When compromise is needed \\[0.1cm]
	\end{tabular*}
	\caption{The Saaty Rating Scale \cite{Coyle.2004}}
	\label{Tab:Saaty}
	\end{adjustwidth}
	\end{table}


\nolinenumbers

\newpage
%
%
%

\bibliography{references.bib}


\end{document}